\documentclass[11pt,a4paper]{article} 
\pdfoutput=1
\usepackage{jheppub}

\usepackage{graphicx} 
\usepackage{amsmath}
\usepackage{xcolor}
\usepackage{booktabs}
\usepackage[normalem]{ulem}

\definecolor{cc}{RGB}{200,80,140}

\usepackage{cleveref}
\crefname{table}{Table}{Tables}
\crefname{equation}{Eq.}{Eqs.}
\crefname{appendix}{App.}{Apps.}
\crefname{section}{Sec.}{Secs.}
\crefname{figure}{Fig.}{Figs.}

\newcommand{\gs}{\ensuremath{\tilde{g}_\star}}
\newcommand{\gss}{\ensuremath{\tilde{g}^2_\star}}
\newcommand{\ratG}{\left(\frac{\gss}{g^2}\right)}

\title{Zeroing in on BSM at Colliders}

\author[a]{Cari Cesarotti,}

\author[a]{Matthew McCullough}

\author[a,b]{and Marc Riembau}

\affiliation[a]{Theoretical Physics Department, CERN, 1211 Geneva 23, Switzerland}
\affiliation[b]{Theoretical Particle Physics Laboratory, EPFL, 1015 Lausanne, Switzerland}

\abstract{
Destructive interference between light and heavy dynamics can produce a zero in the amplitude which, for strongly coupled UV completions, lies parametrically below the heavy threshold. 
We identify under which assumptions this parametrically light zero exists, and when it admits an effective field theory description. We study how absorptive effects and helicity structure affect the existence of the zero and turn it into a dip.  When the dip is kinematically accessible, we show how its position, depth, and curvature constrain the UV dynamics. Collider examples illustrate the importance of interference in indirect searches and the information available when the zero is kinematically accessible. 
}

\begin{document}
\preprint{CERN-TH-2026-285}

\maketitle
\flushbottom

\section{Introduction}
\label{sec:introduction}

The next few decades of high energy collider physics can be broadly characterized as an era of precision.  
Our upcoming collider, the HL-LHC, is an upgrade in statistics rather than energy \cite{ZurbanoFernandez:2020cco}.
The most likely subsequent collider is then FCC-ee (or another EW and Higgs factory), which would fall even more obviously into the precision paradigm \cite{FCC:2018byv, FCC:2018evy, FCC:2025lpp}.
It is therefore imperative to understand how we can use precision machines to not only better our understanding of the Standard Model (SM), but also reveal signs of new dynamics. 

Canonically, there are two limits in which we search for physics beyond the Standard Model (BSM).
With energy-frontier experiments, the parameter space is directly accessible and new physics can be produced on shell.
At the low-energy frontier, precision measurements extend sensitivity to heavy physics through an effective field theory (EFT) description \cite{Contino:2016jqw}.
However, in the intermediate regime of $\sqrt{s} \lesssim \Lambda_\text{NP}$, it can be more difficult to identify new phenomena.
At these energies, new physics can potentially be identified by deficits rather than peaks due to interference effects \cite{Gaemers:1984sj, Bella:2010sc, Bai:2014fkl, Fiaschi:2021sin, BaronMoreno:2026pjr}.

Contrary to na\"ive perceptions, BSM-SM interference effects are not necessarily small \cite{Frixione:2020hqz}.
Even when new physics is weakly coupled, the interference can give an $\mathcal{O}(1)$ correction to the measured rate.
In this work, we quantify when strong interference effects occur and when they can be observed experimentally. 
In simple scattering models, the SM amplitude can be exactly cancelled by the BSM amplitude at a particular energy - we call this phenomenon an \textit{amplitude zero}.
Depending on the parameters of the theory, this can occur at an energy parametrically below the mass of the new particle, making it a promising signature for discovery even at low-energy colliders. 
This possibility has received relatively little attention as a probe of heavy new physics, and we aim to clarify the conditions under which it arises. 
In more realistic settings, this simple picture becomes more complicated, as additional physical effects can shift or lift the zero.
We examine how these effects impact its experimental observability.

Interference between heavy resonances and the SM has been studied in many contexts.
These include heavy scalars decaying to top quarks \cite{Barcelo:2010bm,Bernreuther:2015fts,Hespel:2016qaf,Carena:2016npr,Djouadi:2016ack}, to diboson final states \cite{Maina:2015ela,Kauer:2015hia,Ballestrero:2015jca,Greiner:2016awe,Dawson:2015haa} and to diphoton or digluon final states \cite{Jung:2015etr,Craig:2016ygr,Martin:2016bgw,Bhattiprolu:2020zoq}, as well as heavy vectors \cite{Rizzo:2009pu,Accomando:2013sfa,Accomando:2016mvz,Baur:2007ck,Ferrario:2008wm,Barcelo:2011vk,Choudhury:2011cg,Dasgupta:2023zrh} and higher-spin resonances \cite{Mathews:2004xp,Kumar:2009nn}.
Most of these works focused on the direct discovery of a resonance, where interference distorts the line shape near the peak.
We instead work at the precision frontier and assume the resonance is well beyond kinematic reach.

The remainder of this work is organized as follows.
In Sec.~\ref{sec:zero}, we introduce the basic mechanism underlying amplitude zeros and discuss physical effects that can shift, lift, or obscure their experimental signatures.
In Sec.~\ref{sec:generalisations}, we develop this discussion in more general settings, considering toy models to illustrate how robust the zeros are to physical effects.
In Sec.~\ref{sec:pheno}, we investigate the phenomenological consequences of interference in collider settings, namely at the LHC and FCC-ee.
Through these case studies, we examine how neglecting interference can distort the interpretation of collider data and affect observable distributions and their EFT description. 
Ultimately, we show how interference can provide sensitivity to heavy new physics below the resonance scale.
We conclude in Sec.~\ref{sec:conclusions}.
%


\section{Zeros from Interference}
\label{sec:zero}

\subsection{Basics of Amplitude Zeros}
We begin by laying out a toy scenario to demonstrate the analytic properties of the amplitude zeros. 
Consider the process $\ell^+\ell^-\to \ell^{\prime +}\ell^{\prime -}$, with $\ell\neq \ell^\prime$ as a simplified discussion of SM lepton scattering with new physics. 
We assume that this process, which occurs only through $s$-channel exchange, consists of a massless mediator with coupling $g$, plus a single resonance of mass $m_\star$, coupling $g_\star$.

In the high-energy limit, we can write the amplitude as simply
\begin{align}\label{eq:toyamplitude}
    \mathcal M(s) & = q_1 q_2 \left(\frac{g^2}{s} + \frac{\gss}{s-m_\star^2}\right)(\bar v_\ell \gamma^\mu u_\ell)(\bar u_{\ell^\prime}\gamma^\mu v_{\ell^\prime}) \\
    & \equiv q_1 q_2 g^2s^{-1}A(s)(\bar v_\ell \gamma^\mu u_\ell)(\bar u_{\ell^\prime}\gamma^\mu v_{\ell^\prime}) 
\label{eq:toyamplitude2}
\end{align}
where $\gs^2 \equiv \frac{{q_1}_\star {q_2}_\star}{q_1 q_2} g_\star^2$ is the coupling modified to account for the ratio of charges. 
The first term in Eq.~\ref{eq:toyamplitude} represents scattering via the massless mediator, with coupling $g$, while the second term represents the channel mediated via a heavy vector with mass $m_\star$ and coupling $g_\star$. 
We include this definition of the function $A(s)$ in Eq.~\ref{eq:toyamplitude2} as it encodes the information about interference. 

\begin{figure}
    \centering
    \includegraphics[width=\linewidth]{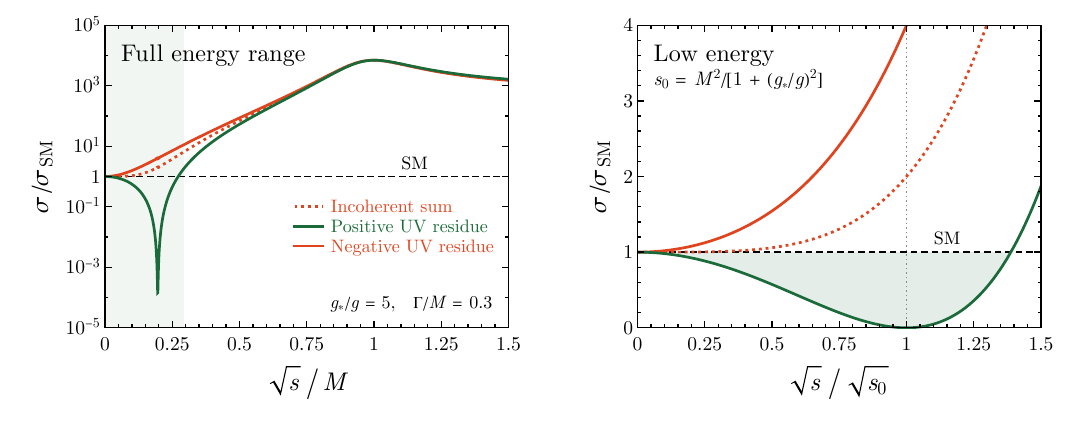}
    \caption{Illustration of the strong interference effect in the (left) full and (right) low-energy ranges. The shaded green region on the left indicates the range of the plot on the right. Positivity implies destructive interference, generating a zero in the amplitude at energies parametrically below threshold.}
    \label{fig:intro_interference}
\end{figure}

Both residues are positive-definite if the charges are of the same sign ($\text{Sign} (\tilde g^2_\star) = \text{Sign} (g^2)$) and, given that they are simple poles, it is guaranteed that the tree amplitude vanishes at some point $s_0$ in between, i.e. $\mathcal{M}(s_0)=0$ for $0 < s_0 < m_\star^2$. 
We solve for this point, the \textit{amplitude zero}, and find it is located at
\begin{equation}\label{eq:ampzero}
    s_0 \,=\, \frac{m_\star^2}{1+\gss/g^2} \, .
\end{equation}
Note that not only is the zero strictly below the resonance, but that for a comparatively strongly coupled resonance for which $\tilde g_\star^2/g^2\gg 1$, the zero appears \textit{parametrically} below threshold, as is illustrated in Fig.~\ref{fig:intro_interference} for the case $g_\star / g= 5$.\footnote{The same result is found in \cite{Bai:2014fkl,Calisto:2026pvv}.} 
This implies that in this toy example, a zero in the amplitude is a distinctive feature accessible to observers at a parametrically lower energy than the direct production of the resonance. 

We mentioned earlier that the existence of the zero is related to the sign of UV dynamics. 
The zero occurring before the pole ($s_0 < m_\star^2$) is a consequence of the amplitude being a sum of simple poles with positive residues.
If the UV residue is negative however, which is possible in many models for distinct initial and final states, the interference is constructive below the pole, and the zero is lifted.
This is also shown in Fig.~\ref{fig:intro_interference}, where the same choice of $(m_\star, g_\star)$ is compared for different signs of the UV residue. 

It is worth noting that positivity, namely large destructive interference, does not produce the same magnitude of deviation in the rate as negative residues for the same choice of parameters. 
At leading order in the derivative expansion, both deviations coincide in absolute value.
However, as energies get closer to the zero $s_0$, destructive interference leads to a $-100\%$ modification of the cross section, while constructive interference yields an enhancement by a factor 4, as visualized in Fig.~\ref{fig:intro_interference}.

\subsection{Complex Zeros}
In the previous section, our discussion of the theory is purely real, which simplifies the prediction of physical zeros. 
However, several physical effects can shift the zero into the complex plane, modifying the analytic properties of the amplitude.

For example, consider how the amplitude is modified when including a width.
Close to resonance, a sufficiently weakly coupled mediator is well-described by a Breit-Wigner, 
\begin{equation}\label{eq:amplitudedeepwidth}
    A(s) \,=\, 1 + \frac{\gss}{g^2} \frac{s}{s-m_\star^2+i \text{Im}\,\Sigma(s)}\,,
\end{equation}
where $\text{Im}\,\Sigma(s)$ is the imaginary part of the self energy. 
We assume that the mediator is sufficiently weakly coupled to fermions and define $\gamma_\star\equiv \Gamma_\star/m_\star$, such that $\text{Im}\Sigma(s) = s\, \gamma_\star$, with $\gamma_\star$ a constant similar to the SM $Z$-boson width. 
This scaling with $s$ softens the effect of the width at low energies. 
The mass, width, and couplings determine the complex locations of both the pole and zero, which are therefore not independent.
Importantly, as the amplitude zero is now off the real axis, we will no longer see a true zero at a physical energy. 
Instead, we observe a dip in the amplitude, but the interference never yields exact cancellation.
This is an important consequence for any experimental search. 
Width effects therefore matter even well below the resonance, where the narrow-width approximation (NWA) is usually assumed to be safe (see e.g.~Ref.~\cite{Berdine:2007uv} for other circumstances in which this approximation fails).

To see how radiative corrections, including finite-width effects, affect this cancellation, we may write
$A(s)=A^{(0)}(s)+\delta A(s)$, where $A^{(0)}$ is the complete tree-level amplitude without width effects and $\delta A$ denotes its leading virtual correction.
For the real tree-level amplitude considered here, suppose that $A^{(0)}(s_0) = 0$.
Assuming that the correction is small near $s_0$, the leading non-vanishing terms near the cancellation give
\begin{equation}
|A(s)|^2 \simeq
\left[
A^{(0)\prime}(s_0)(s-s_0) + \operatorname{Re}\delta A(s_0) \right]^2 + \left[ \operatorname{Im}\delta A(s_0) \right]^2\,.
\end{equation}
The real part of the correction shifts the position of the minimum, while a non-zero imaginary part leaves a non-vanishing minimum in this approximation.
Absorptive contributions arise when intermediate states can go on shell.
These states can be light and appear much earlier than the heavy resonance.
Radiative corrections can therefore prevent the simultaneous cancellation of the real and imaginary parts at a physical energy.
Corrections that multiply both interfering amplitudes by the same finite factor preserve their cancellation, so the lifting of the zero depends on how the corrections modify the relative amplitudes.

An accurate treatment of radiative corrections is also necessary when interpreting a measured suppression of the rate.
At high energies, electroweak Sudakov corrections can substantially suppress SM rates relative to their lowest-order predictions \cite{Chiesa:2013yma,Alioli:2016fum,Denner:2019vbn}.
If these effects are omitted or modeled inadequately, a deficit relative to the assumed SM prediction could be misinterpreted as evidence for destructive SM--BSM interference.
Analyses targeting amplitude zeros must therefore account for the effects that modify or obscure their signatures in measured distributions.

\subsection{Helicity Structure and Amplitude Zeros}
\label{sec:helicityzeros}

The helicity structure of the scattering process is central to determining the strength of interference.
Thus far, we have seen how a parametrically light zero is a generic feature of heavy physics at strong coupling.
However, an exact cancellation for a single helicity amplitude may not occur at the same energy for all helicities. 
This is well known in EFT studies where, in order to obtain the optimal sensitivity, it is necessary to design observables targeting the relevant helicity amplitude, see e.g. Refs.~\cite{Franceschini:2017xkh,Panico:2017frx}. 

The same is true for amplitude zeros, and the SM itself provides an example.
For massless leptons, the $\ell^+\ell^-\to \ell'^+\ell'^-$ (where $l\neq l'$) amplitude splits into four helicity channels, labeled by the chiralities $\chi,\chi'=L,R$ of the two fermion currents.
Each channel receives an $s$-channel photon and $Z$ contribution,
\begin{equation}
  \mathcal{A}_{\chi\chi'}(s) = \frac{e^2}{s}
  + \frac{g^l_{\chi}\, g^{l'}_{\chi'}}{s - m_Z^2}\
  \label{eq:Ahel}
\end{equation}
with $g^\ell_L = \frac{g}{ c_W}\left(-\tfrac12 + s_W^2\right)$ and
$g^\ell_R = \frac{g}{c_W}\, s_W^2$, neglecting the $Z$ width.
The four channels do not interfere with one another and contribute incoherently to the  cross section:
\begin{equation}
  \frac{d\sigma}{d\cos\theta} \propto
  \Big(|\mathcal{A}_{LL}|^2 + |\mathcal{A}_{RR}|^2\Big)(1+\cos\theta)^2
  + \Big(|\mathcal{A}_{LR}|^2 + |\mathcal{A}_{RL}|^2\Big)(1-\cos\theta)^2\,,
  \label{eq:angular}
\end{equation}
where $\theta$ is the angle between the incoming $\ell^-$ and the outgoing $\ell'^-$, and the angular dependence is fixed by the square of the Wigner-$d$ matrices $d^{J=1}_{1, \pm 1}$.
Since $g^\ell_L<0<g^\ell_R$, the $Z$ coupling product $g^{\ell}_\chi g^{\ell'}_{\chi'}$ is positive
for same-chirality channels and negative for opposite-chirality ones, thus shifting the location of the amplitude zero. 

Solving $\mathcal{A}_{\chi\chi'}(s_0)=0$ for each chirality yields
\begin{equation}
  s_0^{\chi\chi'} = \frac{e^2}{e^2 + g^\ell_\chi g^{\ell'}_{\chi'}}\; m_Z^2.
  \label{eq:s0SM}
\end{equation}
From this we see that for the same chirality channel the zero happens before the $Z$ pole, whereas for the opposite chirality it occurs after. 
Thus, in the presence of contributions from multiple chirality configurations the zeros of the SM helicity amplitudes do not coincide and the inclusive rate remains nonzero. 
However, by using the angular information at $s = s_0$, one can still identify significant destructive interference in specific angular bins. 

While the total cross section is proportional to the sum of these terms, the forward-backward asymmetry is proportional to the difference. 
Therefore, the angular dependence of the scattering allows us to separate the same-chirality channels, $|A_{LL}|^2+|A_{RR}|^2$, from the mixed-chirality channels, $|A_{LR}|^2+|A_{RL}|^2$.
A zero, or at least large destructive interference effects, may still be observed by looking at a 2d distribution in both energy and angle. 

\subsection{Effective field theory description}
Finally, we want to highlight that amplitude zeros are compatible with an effective field theory (EFT) description.
An EFT formulation does not require the leading BSM contribution to be small compared with the SM amplitude.
An amplitude zero can therefore arise from the cancellation of two comparable contributions even when higher dimension operators are omitted.

This can be seen directly in the single-resonance example we've already been considering.
The function $A(s)$ of Eq.~\ref{eq:toyamplitude2} can be written as an expansion
\begin{equation}
A(s) = 1- \ratG \frac{s/m_\star^2}{1-s/m_\star^2} = 1- \ratG \left(\frac{s}{m_\star^2} +  \left( \frac{s}{m_\star^2} \right)^2 + \cdots \right) ,
\end{equation}
where $s/m_\star^2 < 1$.
The zero for the dimension-6 approximation $A_{\mathrm{EFT}}^{(6)}(s)=1- \tfrac{\gss}{g^2}\frac{s}{m_\star^2} $ is at
\begin{equation}
s_0^{(6)}=\frac{m_\star^2}{(\gss/g^2)} ~.
\end{equation}
For $\gss/g^2 \gg 1$ this prediction closely approximates the exact zero in Eq.~\ref{eq:ampzero}, which is also when the zero is parametrically smaller than the mass scale. 
The coupling hierarchy required for this separation is relative to the SM coupling, and does not itself require a non-perturbative or large value of $g_\star$.

What is important, however, is retaining the squared terms which might naively be discarded when computing the cross section near the cancellation. 
Squaring the leading EFT amplitude gives
\begin{equation}
\left|A_{\mathrm{EFT}}^{(6)}(s)\right|^2 = 1-2\ratG \left(\frac{s}{m_\star^2}\right)+\ratG^2\left(\frac{s}{m_\star^2}\right)^2.
\end{equation}
Near the cancellation, $\tfrac{\gss}{g^2}\frac{s}{m_\star^2}\simeq1$ such that $A_{\mathrm{EFT}}^{(6)}(s) \simeq 0$, and all three terms are comparable.
The squared dimension-6 contribution must also be retained.
This suggests dimension-8 effects are also relevant, as interference between the SM and dimension-8 amplitudes contributes at the same order as the squared dimension-6 term.\footnote{More details on this power counting can be found in Ref.~\cite{Contino:2016jqw}.}
In this model, however, the dimension-8 contribution is suppressed by the coupling hierarchy, as that contribution includes an extra factor of $g^2/\gss$.

While an EFT truncation may predict the approximate position of the zero reliably, additional terms may be required to describe the rate in the vicinity of the zero.
The utility of a certain-order EFT depends on the scale separation, the coupling ratio, and the precision required of the observable.
We illustrate this in the following phenomenological studies, as tests of particular finite-order descriptions at the energies and precision under consideration.

The EFT description also accommodates the absorptive effects that can turn a tree-level zero into a finite dip, reproducing the effect of the UV particle width.
The light-particle intermediate states that the UV state can decay into remain dynamical after the heavy mediator is integrated out, and their absorptive contributions are reproduced by loops involving the corresponding EFT interactions.
We discuss the relation between the width contribution, the EFT description, and the amplitude zeros in greater depth in Appendix~\ref{sec:width}.


\section{Generalizations}
\label{sec:generalisations}
In the preceding discussion, we established the features of a generic theory necessary to generate amplitude zeros, as well as identified effects that can mask where amplitudes cancel.
We now examine some phenomenological consequences and how we might construct observables to understand the origin of interference effects even when a full zero is not visible.

\subsection{Chiral Couplings and a Heavy Mediator}
\label{sec:chiralcouplingsgeneral}

To begin, we again consider a toy model which mimics the structure of the SM interfering with a new heavy mediator. 
We include the $Z$ in our toy to emphasize the effect of chiral couplings.
We can generalize this discussion, assuming now that the IR amplitude is the one given by the SM photon and the $Z$-boson channels, while a heavy vector couples to the SM fermions in a generic way. 
Considering $\ell^+\ell^-\to \ell^{\prime +}\ell^{\prime -}$ scattering, keeping track of the chiralities of the $\ell$- and $\ell^\prime$-currents, one can write $A(s)$ as
\begin{eqnarray}\label{eq:amplitudezerochiral}
    A_{\chi \chi^\prime}(s) &=& e^2 +  \frac{g^\ell_\chi g^{\ell^\prime}_{\chi^\prime} s}{s-m_Z^2+i s \gamma_Z} +  \frac{g^\ell_{\star\chi}g^{\ell^\prime}_{\star \chi^\prime} s}{s-m_\star^2+i s \gamma_\star} ~~,
\end{eqnarray}
where the $e^2$ denotes the photon coupling, independent of the chirality.
In the limit $m_Z^2 \ll s_0$, we can write the chirality-dependent amplitude zero as
\begin{equation}
s_0 = \frac{m_\star^2}{ 1+ \frac{g^\ell_{\star\chi}g^{\ell^\prime}_{\star\chi^\prime}}{e^2+g^\ell_\chi g^{\ell^\prime}_{\chi^\prime}} } ~~.
\end{equation}
Thus, existence of the zero below threshold depends on the sign of the SM coupling combination.
Denoting $g^{2,SM}_{\chi \chi'} = e^2+g^\ell_\chi g^{\ell^\prime}_{\chi^\prime}$ we find that in the SM and for leptons, not only are the diagonal terms positive, $g^{2,SM}_{LL} \simeq 0.14$,  $g^{2,SM}_{RR} \simeq 0.13$, but also the off-diagonal $g^{2,SM}_{RL}=g^{2,SM}_{LR} \simeq 0.06$. 
The zero then depends on the UV couplings, which we assume to be equal for any lepton, and we parametrize in terms of $g_\star$ and $\theta_\text{UV}$ as $g_{\star L} = g_\star \cos\theta_\text{UV}$ and $g_{\star R} = g_\star \sin\theta_\text{UV}$. 

Note that $\theta_\text{UV}\to \theta_\text{UV}+\pi$ changes the sign of both $g_{\star L}$ and $g_{\star R}$ and therefore leaves the residues and thus the squared amplitudes invariant, so we can consider $\theta_\text{UV}\in [-\frac{\pi}{2},\frac{\pi}{2}]$. 
Within this range, each chirality configuration has further redundancies. 
For $\theta_\text{UV}\to -\theta_\text{UV}$, $g_{\star L,R}\to \pm g_{\star L,R}$, and therefore the same-chirality term is invariant, so the independent range can be chosen to be $\theta_\text{UV}\in [0,\pi/2]$. 
Instead, for $\theta_\text{UV}\to \pi/2-\theta_\text{UV}$, one has $g_{\star L}\leftrightarrow g_{\star R}$, and it leaves invariant only the mixed-chirality process. 
Thus, the independent range can be chosen to be $\theta_\text{UV}\in [-\pi/4,\pi/4]$. A combined measurement may disentangle the different regions. 

\begin{figure}
    \centering
    \includegraphics[width=\linewidth]{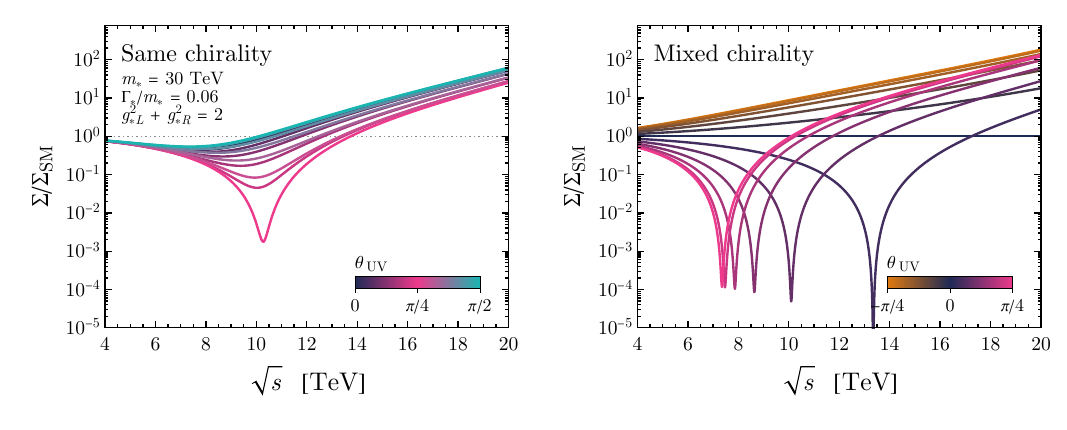}
    \caption{Total rates for the same-chirality channel, $|A_{LL}|^2+|A_{RR}|^2$, and mixed-chirality channel, $|A_{LR}|^2+|A_{RL}|^2$, as a function of the center of mass energy $\sqrt{s}$ and for the point $m_\star=30\,\text{TeV}$, $\gamma_\star=0.06$ and $g_\star^2=2$.}
    \label{fig:chiralzeros}
\end{figure}

For the same-chirality processes, given that the SM couplings are relatively similar, the cancellation will coincide in both amplitudes at the point $\theta_\text{UV}\simeq \pi/4$. 
As $\theta_\text{UV}\to 0,\pi/2$, i.e.\ the purely chiral coupling, the resonance may cancel one of the amplitudes, but it decouples from either the left or the right channel, and therefore the zero of the combined amplitude is lost. 
In these limits the zero becomes a decrease in the rate by a factor $\sim 2$. 
Note that due to positivity of the couplings, the interference is always negative. 
This is illustrated in the first panel of Fig.~\ref{fig:chiralzeros}, where we show the same-chirality rate for a choice of parameters $m_\star=30\,\text{TeV}$ and $g_\star^2=2$. 
Assuming a decay to only the two lepton modes, this gives $\Gamma_\star/m_\star\simeq 6\%$. 

Instead, for the mixed-chirality channel, universality of the couplings implies that $|A_{LR}|=|A_{RL}|$, and therefore there is a single channel. 
However, since now positivity of the residue is not guaranteed, the existence or not of a zero depends on the sign of the couplings. 
For $\theta_{UV}$ negative, the residue is negative, and therefore the interference is constructive. 
In this regime interference enhances the rate and there is no zero. 
The zero of the amplitude then exists for positive values of $\theta_\text{UV}$, as is illustrated in Fig.~\ref{fig:chiralzeros}.


\subsection{Zeros from a spectral density}
\label{sec:strongcoupling}
The possibility of amplitude zeros isn't restricted to perturbative UV-completions. 
The same set of assumptions that go in the arguments for the presence of positivity in a given scattering channel, as argued in \cite{McCullough:2023szd}, can be used to derive the existence of a zero at energies parametrically lower than the first branch cut of the theory.
For instance, consider the scattering of some light states. 
A bilinear current of these states excites an operator $\mathcal O$ of some weakly or strongly coupled sector:
If the scattering is through a single channel, using the non-perturbative K\"all\'en-Lehmann representation for the two-point correlator of $\mathcal O$, we may write the UV contribution to the amplitude by inserting the two-point function for the heavy new states.

By extracting the common product of external currents and normalizing to the IR amplitude, one obtains $A(s)$,
\begin{equation}\label{eq:ampspectraldensity}
    A(s)\,=\, 1 + \frac{g_\star^2}{g^2} s R(s)\,\,,\hspace{1cm}\text{with}\hspace{0.5cm}
    R(s)=\int_{m^2_\text{th}}^\infty \frac{d\mu}{\mu} \frac{\rho(\mu)}{s-\mu}\,,
\end{equation}
where the density of states $\rho(\mu)\geq 0$ is positive and corresponds to the spectral density of $\mathcal O(x)$, $(2\pi)\theta(p^0)\rho(p^2)=\sum_\alpha |\langle 0|\mathcal O|\alpha\rangle|^2$. 
In the case of having several currents, the spectral density becomes a positive semi-definite matrix.  

We have assumed for simplicity that only $s$-channel contributions exist, which restricts the class of possible UV models for which this description holds, but we have not limited the UV to be necessarily perturbative.
The assumption is that $\rho(\mu)$ scales as a constant as $\mu\to\infty$, which is guaranteed for instance in the case of conserved currents, see \cite{McCullough:2023szd}.  In such cases boundary terms (counter-terms) with undefined magnitude and sign are avoided.

Note that this construction can be used generally. 
For the tree-level exchange of multiple perturbatively-coupled states, the density of states reduces to a sum of Dirac-$\delta$ functions. 
For instance, $\rho(\mu) = m_\star^2 \delta(m_\star^2-\mu)$ leads to the single-resonance model of Eq.~\ref{eq:toyamplitude}. 

Since the derivative of $A(s)$ is negative, as  
\begin{equation}
  A'(s) \sim R(s)+sR^\prime(s)=-\int_{m^2_\text{th}}^\infty d\mu \frac{\rho(\mu)}{(s-\mu)^2} < 0 ,
\end{equation}
the zero exists below threshold $m_\text{th}$ as long as the amplitude at threshold obeys $-\frac{g_\star^2}{g^2}s_\text{th} R(s_\text{th})>1$, with $s_\text{th} \lesssim m_\text{th}^2$.
This condition is more easily satisfied when there is a coupling hierarchy $g^2/g_\star^2\ll 1$, but it is not necessary. 
The dispersive contribution near threshold should be sufficiently large to cancel the IR contribution to the amplitude, which is not guaranteed if the dispersive integral receives the bulk of the contribution at $\mu\gg m^2_\text{th}$. 
When the condition is satisfied, the zero obeys
\begin{equation}\label{eq:zerodispform}
    \frac{g_\star^2}{g^2}s_0 \,R(s_0)+1=0\,,
    \quad\text{with}\quad 0<s_0<m_{\text{th}}^2\,.
\end{equation}

At low energies, we can write the amplitude in the form
\begin{equation}\label{eq:sumcn}
    A(s) \,=\, 1 - \frac{g_\star^2}{g^2}\sum_{n\geq 1} c_n \hat{s}^n
\end{equation}
with the hat denoting the dimensionless $\hat{s}=s/m^2_\text{th}$, and 
with $c_n$ being identified with the moments of the spectral density, $c_n=(m^2_\text{th})^n\int_{m^2_\text{th}}^\infty \frac{d\mu}{\mu} \frac{\rho(\mu)}{\mu^n}$, which obey positivity $c_n\geq 0$, monotonicity $c_n\geq c_{n+1}$, \cite{Englert:2019zmt}. 

Finally, we want to highlight that the spectral-function formulation of the amplitude zeros is compatible with an EFT description. 
Taking into account only the leading term in the EFT expansion of the amplitude, 
the zero is predicted to be at $s_0^{(6)} = m_{\rm th}^2(g^2/g_\star^2)/c_1$.
This coincides with Eq.~\ref{eq:ampzero} in the $g^2/g_\star^2 \ll 1$ limit for $c_1=1$, since in this limit the zero is parametrically light and the amplitude there is well described by the leading term in the EFT.

\subsection{Characterizing an accessible dip}
\label{sec:low_eng}
In the single-resonance model, measurements near $s\simeq m_\star^2$ can be used to constrain the mass, coupling and width of the mediator, which jointly determine the amplitude and interference pattern.
When only the low-energy destructive interference dip is accessible, it is useful to characterize the measured distribution directly through the position of the minimum, the rate at that minimum, and the curvature around it.
We therefore define
\begin{equation}
    s_0\,=\,\text{arg}\,\text{min}_{s>0} |A(s)|^2\,,\hspace{1.15cm}
    \gamma^2=|A(s_0)|^2\,,\hspace{1.15cm}
    \mathcal K^2 = \frac{s_0^2}{2}\frac{d^2}{d s^2}|A(s)|_{s = s_0}^2\,,
\end{equation}
namely, the rate normalized to the SM rate $|A(s)|^2$ is determined by the position of the minimum of the dip $s_0$, the depth of the dip and the curvature $\mathcal K^2$ around it. 
Thus, at $s\simeq s_0$, $|A(s)|^2\simeq \gamma^2 + \mathcal K^2 \frac{(s-s_0)^2}{s_0^2}$. 
Within this single-resonance model, these three observables can be mapped one-to-one to the three UV parameters $(m_\star,g_\star,\gamma_\star)$. 
Using the model of Eq.~\ref{eq:amplitudedeepwidth}, and at leading order in $\gamma_\star^2$, one obtains~\footnote{
Keeping finite $\gamma_\star^2$ corrections leads to the expressions
$s_0=\frac{2m_\star^2}{g_\star^2/g^2+2+\Delta}$, $\gamma=\frac{2\gamma_\star}{g_\star^2/g^2+\Delta}$ and $\mathcal K=
\frac{(1+g_\star^2/g^2-\gamma^2)\sqrt{1-\gamma^2}}
     {g_\star^2/g^2\sqrt{1+\gamma^2}}$, 
with $\Delta=\sqrt{(g_\star^2/g^2)^2+4\gamma_\star^2}$.
}
\begin{equation}\label{eq:irtouvmappingsingleresonance}
    s_0 \,=\,m_\star^2 \frac{1}{1+g^2_\star/g^2}\,,\hspace{1.5cm}
    \mathcal K \,=\, 1+\frac{g^2}{g^2_\star}\,,\hspace{1.5cm}
    \gamma \,=\, \gamma_\star\frac{g^2}{g_\star^2}\,.
\end{equation}
Note that $s_0$ is below threshold. 
The curvature is directly a function of the ratio of UV and IR couplings, which with the position of the zero determines the mass of the resonance as 
\begin{equation}\label{eq:mstarfroms0K}
m_\star^2\,=\,s_0\,\frac{\mathcal K}{\mathcal K-1} ~~,
\end{equation}
in the small width limit.   If the depth of the dip $\gamma$ can be measured, then one could determine the width of the UV resonance as well. 

Any model giving rise to a zero will determine the IR observables in terms of some UV parameters. 
We can do so for the real amplitude below the spectral threshold in the representation of Eq.~\ref{eq:ampspectraldensity}, for which the minimum is an exact zero and given implicitly as in Eq.~\ref{eq:zerodispform}, and the curvature $\mathcal K = -s_0A^\prime(s_0)$ by
\begin{equation}
\mathcal K = 1+\frac{g_\star^2}{g^2}s_0^2\int_{m^2_\text{th}}^\infty \frac{d\mu}{\mu} \frac{\rho(\mu)}{(s_0-\mu)^2}~.
\end{equation}
In fact, one can write the previous expression as $\mathcal K=1+\frac{d\ln R}{d\ln s}\vert_{s=s_0}$. 

A low-energy observer, then, via $s_0$ and the curvature can define an effective mass $m_\star$ via Eq.~\ref{eq:mstarfroms0K}, obtaining
\begin{equation}
    m_\star^2 \,=\, \frac{1}{\int_{m^2_\text{th}}^\infty \frac{d\mu}{\mu} \frac{\rho(\mu)}{(s_0-\mu)^2}}\int_{m^2_\text{th}}^\infty \frac{d\mu}{\mu} \frac{\mu \,\rho(\mu)}{(s_0-\mu)^2}\,,
\end{equation}
namely a low energy observer would measure an effective mass $m_\star^2$ corresponding to the first moment of the measure $\frac{d\mu}{\mu} \frac{\rho(\mu)}{(s_0-\mu)^2}$.
Given it is a positive average, one obtains $m_\star^2\geq m^2_\text{th}$, so the effective mass $m_\star^2$ is an upper bound on the threshold of new physics. 

Using the form in terms of moments of Eq.~\ref{eq:sumcn}, we get that the curvature is given by 
$\mathcal K = \frac{g_\star^2}{g^2}\sum_{n\geq 1} n c_n \hat s_0^n$, where the hat denotes a normalization by the UV threshold, $\hat s=s/m^2_\text{th}$, and with $\hat s_0$ being the position of the zero. 
The condition for the zero at $\hat s_0$ can be written now as $\frac{g_\star^2}{g^2}\sum_{n\geq 1}c_n \hat s_0^n=1$. 
Inserting this constraint to the definition of the curvature, one has the equivalent representation $\mathcal K = 1+\frac{g_\star^2}{g^2}\sum_{n\geq 1} (n-1) c_n \hat s_0^n$, with the inequality saturated for a pure dimension 6, or $n=1$, contribution. 
For a zero below threshold, $\hat s_0<1$ and the measurement of the curvature sets an upper bound on the cutoff of the theory, given by
\begin{equation}
    m^2_\text{th} \, \leq \, s_0 \frac{\mathcal K}{\mathcal K -1}\,.
    \label{eq:lowerLambdabound}
\end{equation}
At strong coupling $g^2/\gss\ll 1$, one recovers the $g_\star^2/g^2$ gap between the zero and the resonance.
However, Eq.~\ref{eq:lowerLambdabound} is equivalent to the expression obtained before, and it sets an upper bound on the cutoff only from the low energy observable $\mathcal K$ and the position of the zero $s_0$, and in particular it is independent of $g_\star$ or the EFT expansion convergence. 

The curvature carries information of the relative importance between the dimension 6 term squared and the dimension 8 whenever these are the two dominating contributions from the EFT expansion. Given that $\mathcal K=-s_0A^\prime(s_0)$, one gets
\begin{equation}
    \mathcal K \,\simeq\, 1 + \frac{g^2}{g_\star^2}\frac{c_8}{c_6^2}\,.
\end{equation}
Note that this argument only involves the amplitude and is independent of measuring linear or quadratic effects in the amplitude squared. However, this information is in $\frac{g^2}{g_\star^2}$-suppressed effects. 

Thus, while a resonance is described by its position, total rate and width, a near-zero of the amplitude due to strong destructive interference of UV dynamics is described by the position of the dip $s_0$, its curvature $\mathcal K$, and the depth $\gamma$. These observables exist independently of the convergence of the EFT expansion and can be used to determine properties of the UV dynamics, as illustrated in Sec.~\ref{sec:zeroskinematic}.

\section{Interference at Colliders}
\label{sec:pheno}
In this section, we study how zeros and interference affect sensitivity to heavy new physics in current and future collider settings.
We examine effects on several models of new physics which could not only improve existing LHC analyses, but also aid to make potential discoveries at the FCC-ee or a muon collider. 
Particular emphasis is placed on sensitivity below the heavy-particle mass scale, where interference can modify measured distributions even when no distinct dip is resolved.
We also explore how precision measurements can constrain the energy scale of an amplitude zero, and how an accessible dip could be characterized at a future muon collider to obtain information about the underlying heavy dynamics.

\subsection{LHC interference effects: the $W'$ example}

In order to demonstrate the phenomenological consequences of amplitude zeros we begin with the example of a sequential $W'$ boson at the LHC.
There are a few reasons for this choice.
The first is that $W'$ bosons arise in many motivated BSM scenarios, including composite Higgs and extra-dimensional models.
The second is that for the example of leptonic decays, due to the missing energy in the final state, it is not a scenario in which one may easily anticipate the impact of an amplitude zero.
Thus we use this example to highlight that the effect can be significant, even if the zero is not visible in terms of a collider observable.
Thirdly, there already exist ATLAS and CMS $W'$ searches, hence the aspects raised here may find immediate application.

Our benchmark here is motivated by the CMS analysis \cite{CMS:2022krd}.\footnote{We note that a comparable ATLAS analysis \cite{ATLAS:2019lsy} has also been presented.  We only study the CMS analysis here for the sake of brevity.}
In Ref.~\cite{CMS:2022krd} a search for a sequential $W'$ was performed, amongst other things, and an expected lower limit on the mass was found to be $M_{W'} > 5.6$ TeV at 95\% confidence level using the $e,\mu+\nu$ Drell-Yan channel.
In Ref.~\cite{CMS:2022krd} and the comparable ATLAS analysis \cite{ATLAS:2019lsy} the amplitude interference was not included, with the SM and $W'$ differential distributions added separately.\footnote{Note that while it was not included in this study, interference in charged-current processes has been previously studied. For example, in the $t\bar b$ final state in Ref.~\cite{Boos:2006xe} and for the $\ell\nu$ final state in Refs.~\cite{Rizzo:2007xs,Papaefstathiou:2009sr}.  Particularly notable is Ref.\ \cite{Accomando:2011eu} concerning $W'$ scenarios.  In this work we emphasize that the large effects are a result of the amplitude zero.}
Thus the same benchmarks provide an opportunity in this work to consider how the reach may be affected, and enhanced, once interference effects are included.

\begin{figure}[t]
\begin{center}
\includegraphics[width=.48\textwidth]{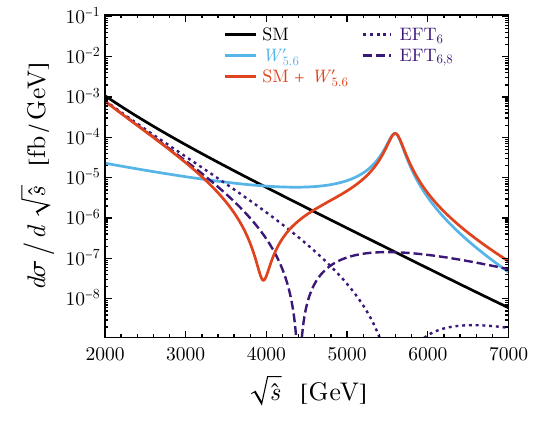} ~ \includegraphics[width=.48\textwidth]{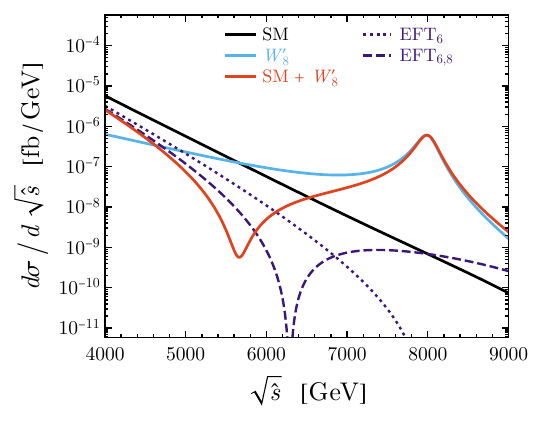} 
\caption{Center of mass distribution of the $l\nu$ final state in the SM (black) alongside $W'$ contributions alone (blue), with interference (red), and with only dimension-6 (dotted purple) and dimension-6+8 (dashed purple) terms included in the BSM amplitude.  The left (right) panel is for a $5.6$ ($8$) TeV sequential $W'$.}
\label{fig:CM}
\end{center}
\end{figure}

In Fig.~\ref{fig:CM} the $l\nu$ final state CM distribution is shown under various BSM assumptions.  Importantly, this distribution is not visible in terms of collider observables, but is shown here to inform later discussion on the impact of zeros. 
Both examples show strong interference and a clear amplitude zero, as well as the efficacy of the EFT description.
Note that at dimension-6 in the EFT description the resonance zero must lie exactly at the resonance mass, due to the equal coupling with the SM, thus it lies strictly beyond EFT validity.  Nonetheless, for the region where the description is valid, below resonance, it provides a reasonably good description of the interference, improving further once dimension-8 is included.  A signal estimate based on a NWA-inspired approach with SM and signal squared and added separately would clearly give a very poor description of the data in the CM regions below $s \lesssim M^2_{W'}/2$, indicating the importance of including interference in the precision era where new resonances may be beyond direct kinematic reach, but within the reach of precision measurements.

\begin{figure}[t]
\begin{center}
\includegraphics[width=.6\textwidth]{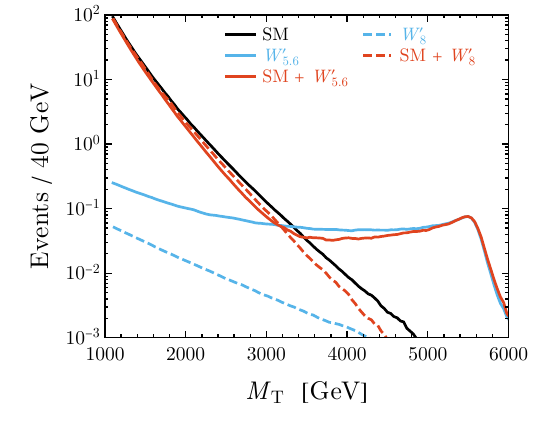} 
\caption{The $M_T$ distribution for the SM, the 5.6 TeV $W'$ alone, and the combined event rate with interference included.  The same is also shown for an 8 TeV $W'$.  At low $M_T$ values the destructive interference following from the high $\sqrt{s}$ region of Fig.~\ref{fig:CM} is apparent.} 
\label{fig:MT}
\end{center}
\end{figure}

In Fig.~\ref{fig:MT} we instead plot  the distribution of events in the $M_T$ variable, which is commonly used for charged current events with a $l\nu$ final state.  For a $5.6$ TeV resonance we see that below $\sim 3$ TeV the interference effect is significant, also for an $8$ TeV resonance.  The reason for this is that many events which populate this $M_T$ region originate from underlying events with greater CM energy, typically a factor $\sqrt{2}$ greater, particularly in and around a zero.  Hence, despite being reasonably far below the mass of the resonance, this $M_T$ window is significantly impacted by the amplitude zero.

\begin{table}[htbp]
  \centering
  \renewcommand{\arraystretch}{1.15}
  \begin{tabular}{lcccc}
    \toprule
    $M_{\mathrm{T}}~(\mathrm{TeV})$
      & $>1.0$ & $>2.0$ & $>3.0$ \\
    \midrule
    Electron data
      & $831$ & $23$ & $1$ \\
    Total SM backgrounds
      & $820$
      & $21$
      & $1.2$\\    
    Uncertainties
      & $\pm 64$
      & $\pm 2.5$
      & $\pm 0.24$\\
    Statistical
      & $\pm 29$
      & $\pm 4.5$
      & ${}^{+0.8}_{-1.2}$\\
    $\delta^{\text{\sout{Int}}}_{5.6~\mathrm{TeV}}$  
      & $9.7$
      & $5.5$
      & $3.6$\\
    $\delta^{\text{Int}}_{5.6~\mathrm{TeV}}$
      & $-88$
      & $-4.5$
      & $2.0$\\
    \bottomrule
  \end{tabular}
  \caption{Observed events in \cite{CMS:2022krd} for the electron final state, alongside statistical errors, combined systematic and statistical errors taken directly from \cite{CMS:2022krd}, and predicted BSM event rate corrections without and with interference in the amplitude zero taken into account.}
  \label{tab:CMScomparison}
\end{table}

These observations are quantified in Tab.~\ref{tab:CMScomparison}.  The SM predictions are similar to those of \cite{CMS:2022krd}, and systematic and statistical errors are taken directly from that reference.  Using the $M_T > 3$ TeV bin we find that the uninterfered $M_{W'}^{\text{\sout{Int}}}$ signal is excluded at the 96\% C.I. consistent with the findings of \cite{CMS:2022krd}.  However, the signal in that bin is reduced in the presence of interference, bringing the discrepancy down to the $\sim 80\%$ C.I. level.  The significant interference at lower energies does bring the $M_T > 1$ TeV bin into a $\gtrsim 1 \sigma$ tension as well, where the sensitivity is significantly impacted by larger systematic errors and a reliable estimate of the revised mass sensitivity would require a detailed likelihood profile.  We estimate that the combined $M_{W'}$ limit may be similar, though possibly mildly weaker, than the one found in \cite{CMS:2022krd}, however it is clear that the amplitude zero significantly impacts the nature of the signal shape and the interplay with background.  Furthermore, were the systematic errors in the lower bin to be significantly reduced it could be that, due to the significant interference, this bin could dominate sensitivity at the HL-LHC.\footnote{Similar conclusions on the importance of including interference were drawn in \cite{Fiaschi:2021sin}.}

\subsection{Interference at FCC-ee}

We now shift our attention to FCC-ee. 
New physics within kinematic reach of FCC-ee that evades existing constraints generally has either very weak couplings overall or particularly suppressed couplings to colour-charged particles.
As shown in Eq.~\ref{eq:ampzero}, a zero will appear parametrically lower than the mass if the coupling is strong, or at least much stronger than the relevant SM coupling ($\tilde{g}_*^2/ g^2 \gg 1$). 
In theory, this means that particles of $\mathcal{O}(1-10)$ TeV could be detected at the $t\bar{t}$ run ($\sqrt{s} = 365$ GeV) at the FCC-ee.

To illustrate the reach of the FCC-ee we consider two new physics scenarios from additional $U(1)$ gauge symmetries, both of which evade LHC constraints by coupling only to the lepton sector at tree level.
The first is a leptophilic $U(1)$, where all the leptons have the same charge and coupling. 
Such a coupling structure can arise from gauging total lepton number, $L=L_e+L_\mu+L_\tau$, with additional fermions introduced to cancel the gauge anomalies \cite{FileviezPerez:2010gw, Schwaller:2013hqa}.
The second is the anomaly-free gauged lepton number symmetry, $L_\mu - L_e$ (see e.g.\ \cite{Dasgupta:2023zrh,GonzalezSuarez:2024dsp}). 
The Lagrangian of both can be summarized as 
\begin{equation}
\mathcal{L} \supset -\frac{1}{4} Z'_{\mu\nu} Z'^{\mu\nu}
+ \frac{1}{2} M_{Z'}^2 Z'_\mu Z'^\mu
+ g_*\, Z'_\mu \sum_{\ell = e,\mu,\tau} Q'_\ell
\left( \bar{\ell}\, \gamma^\mu \ell + \bar{\nu}_\ell\, \gamma^\mu P_L\, \nu_\ell \right) ,
\end{equation}
with the charge assignments
\begin{equation}
\text{leptophilic:}\; Q'_f = +1 , \qquad
L_\mu - L_e:\; Q'_e = +1,\; Q'_\mu = -1,\; Q'_\tau = 0.
\end{equation}
The leptophilic model yields a deficit due to the destructive interference, whereas the $L_\mu - L_e$ model shows a smooth excess well before the peak.
We focus on the $e^+e^- \rightarrow \mu^+ \mu^-$ signature rather than the $e^+e^-$ final state to avoid interfering with the $t$-channel process. 
Our results assume $\sqrt{s} = 365$ GeV with $\mathcal{L}$ = 3.12 ab$^{-1}$ of luminosity.

We consider the differential cross section in both the invariant mass of the muon system $m_{\mu \mu}$ as well as the angle of the muon $\theta_{\mu^-}$.
Both are needed in combination to identify the most significant interference effects, as discussed in Sec.~\ref{sec:zero}.

\begin{figure}[t!]
    \centering
    \includegraphics[width=0.4\linewidth]{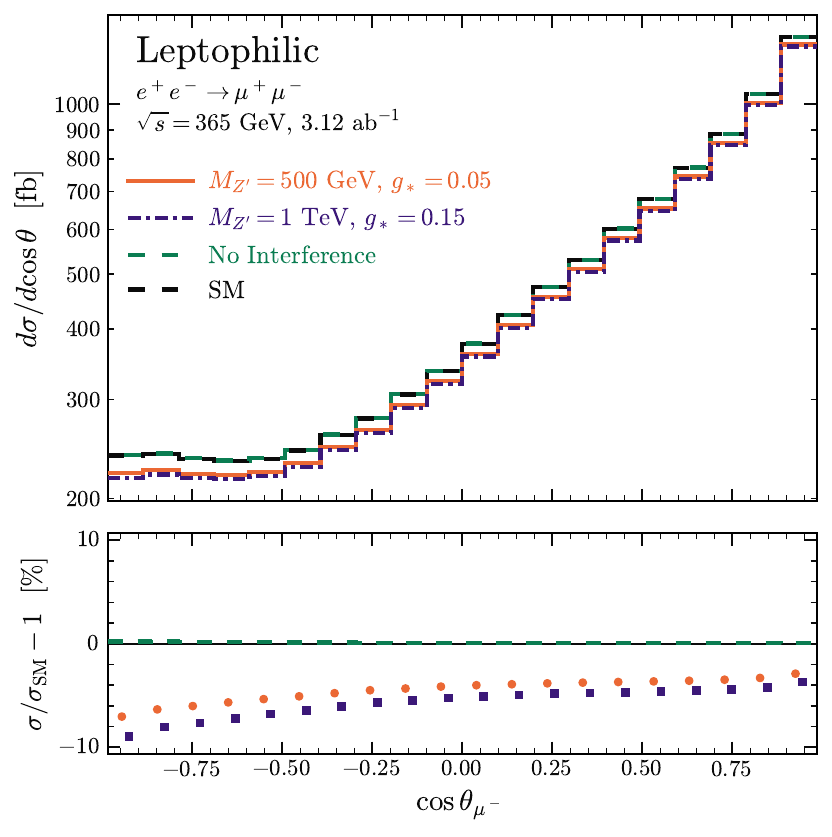} \qquad
    \includegraphics[width=0.4\linewidth]{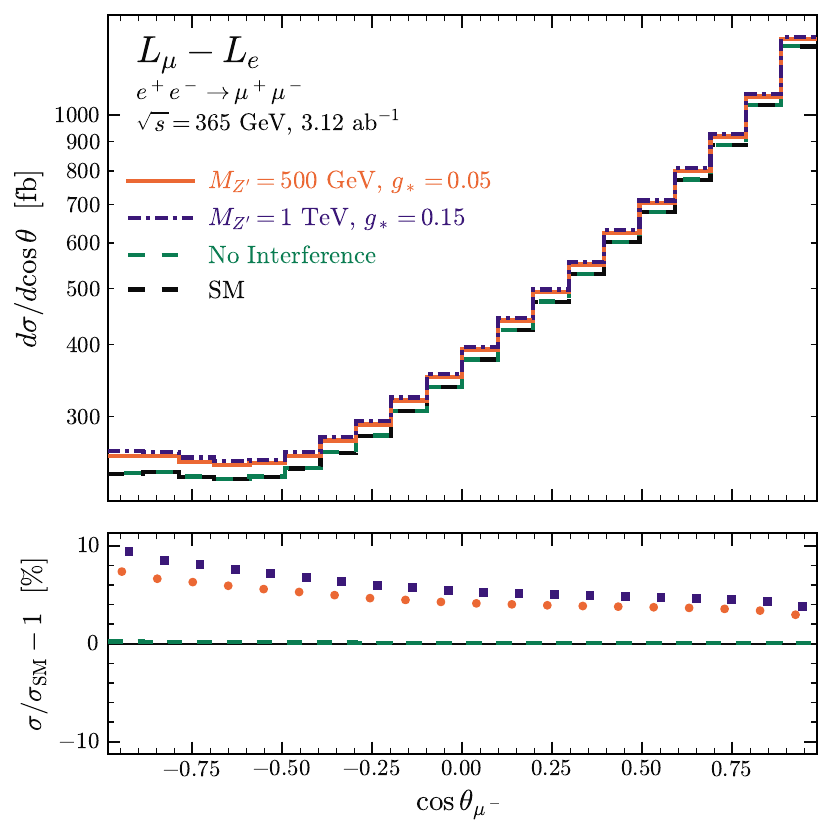}
    \caption{Differential distribution of the cross section for $e^+e^- \rightarrow \mu^+ \mu^-$ for both the leptophilic (left) and $L_\mu - L_e$ (right) models. 
    We show the SM result, as well as the full result including interference with $M_{Z'} = 0.5, 1$ TeV.
   For both benchmark masses, the predictions obtained without interference are nearly indistinguishable from the SM curve.}
    \label{fig:costh_benchmarks}
\end{figure}

Two benchmark masses, $M_{Z'} = 0.5, \ 1$ TeV, are depicted in Fig.~\ref{fig:costh_benchmarks}.
From these plots, it is clear that both destructive and constructive effects are sensitive to the heavy new physics well before the mass pole. 

\begin{figure}[h!]
    \centering
    \includegraphics[width=0.45\linewidth]{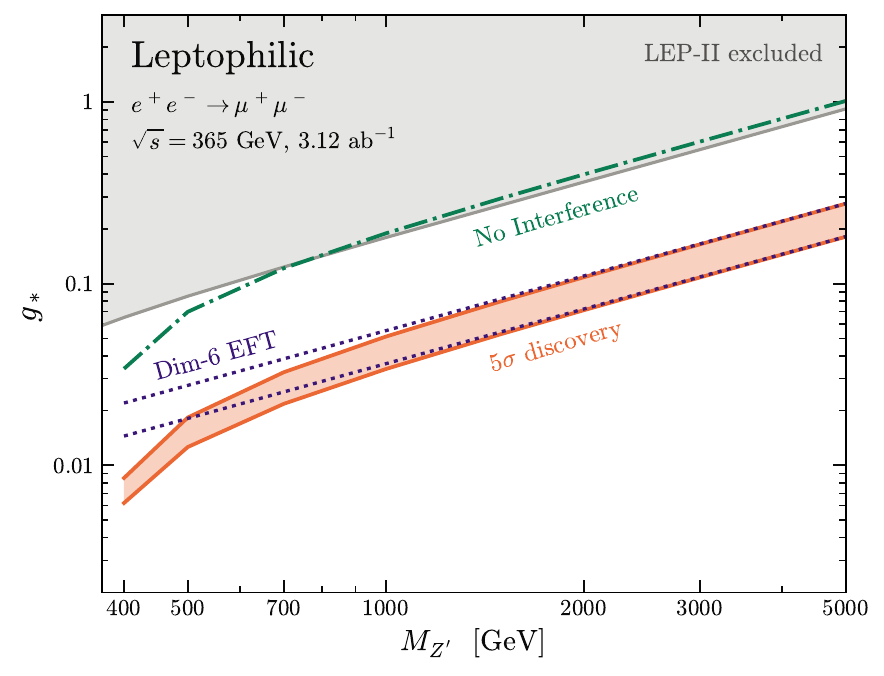}\qquad
    \includegraphics[width=0.45\linewidth]{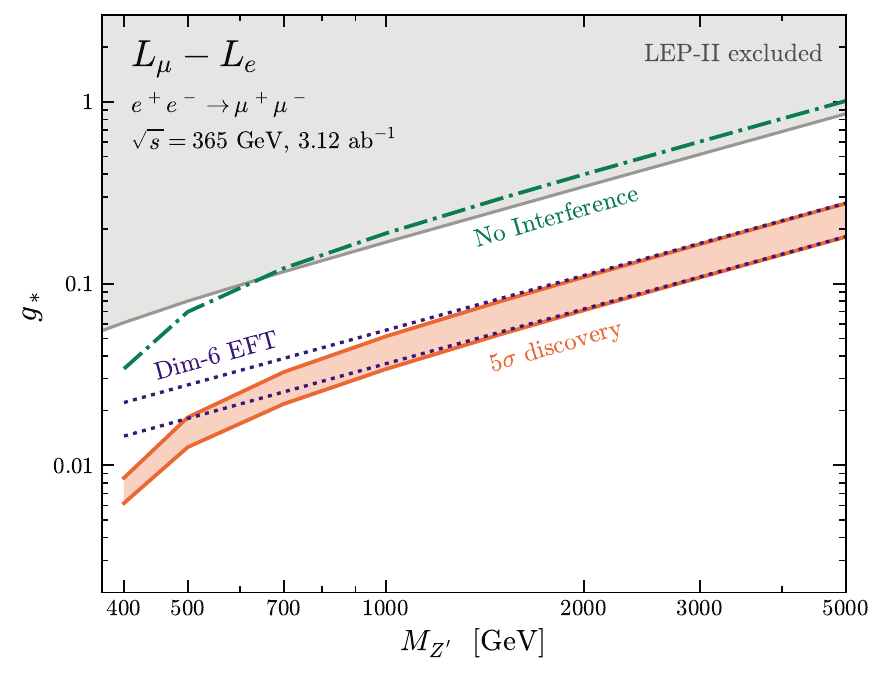}
    \caption{The 5$-\sigma$ discovery plots for the leptophilic (left) and $L_\mu - L_e$ (right) models.  }
    \label{fig:FCC_reach}
\end{figure}

To quantify the reach of FCC-ee, we compute the expected discovery significance with a binned $\chi^2$, 
\begin{equation}
Z^2 = \mathcal{L} \times \sum_i \frac{\left(\sigma_i^\text{SM$ + Z'$} - \sigma_i^\text{SM}\right)^2}{\sigma_i^\text{SM}}
\label{eq:FCC_testStat}
\end{equation}
summed over the bins in $m_{\mu\mu}$ and $\cos\theta$.
With this test statistic, we are equally sensitive to deficits and excesses. 
We then solve for the minimum coupling $g_*$ which gives $Z = 5$. 
The results of this procedure are shown in Fig.~\ref{fig:FCC_reach}. 

As we cannot yet claim the efficiency of detection or acceptance at the future collider, it is possible that the absolute event rate will carry an uncertainty larger than the signal itself. 
This assumes a sub-percent effect on the total number of events per bin.
The discovery band in Fig.~\ref{fig:FCC_reach} therefore spans from a known normalization (lower edge, optimistic reach) to a floating one, where only the shape of the distributions is used (upper edge, pessimistic reach).
Additionally, we consider the reach where the results are incoherent, e.g. $\sigma^\text{SM$ + Z'$} = \sigma^\text{SM}+\sigma^\text{$Z'$}$, to emphasize the importance of interference effects. 
The leading constraints for these models were set by LEP-II \cite{ALEPH:2013dgf}. 
We can also consider how an EFT approach would perform for this search. 
To model the effect with a dimension-6 EFT, we replace the $Z'$ propagator $(\hat{s} - M_{Z'}^2)^{-1} \rightarrow - M_{Z'}^{-2} $ when evaluating the test statistic in Eq.~\ref{eq:FCC_testStat}. 
In this limit, the $Z'$ enters only through the contact coefficient $g_*^2/M_{Z'}^2$, so there is a redundancy on theories with the same coupling-to-mass ratio.
The EFT therefore gives a single bound on $g_*/M_{Z'}$, which we can estimate by 
\begin{equation}
Z^2 \approx \mathcal{L} \times \sum_i \frac{(\sigma^\text{SM}_i - \sigma^\text{EFT}_i)^2}{\sigma_i^\text{SM}} \approx  \mathcal{L} \times \frac{ s g_*^4}{ 3 \pi M_{Z'}^4}.
\end{equation}
This equates to $g_*/M_{Z'} \gtrsim (28~\text{TeV})^{-1}$ at $5\sigma$, corresponding to a straight line in the $(M_{Z'}, g_*)$ plane.
As anticipated by the expansion, the EFT approximation improves as the mass of the $Z'$ increases. 
At $M_{Z'} = 500$~GeV, the EFT underestimates the reach by a factor of
$\sim 1.4$ in $g_*$, while above $M_{Z'} \gtrsim 2$ TeV, they agree to within a few percent.
The difference between the EFT and the full simulation is summarized in Tab.~\ref{tab:eft_vs_full}.

\begin{table}
\centering
\begin{tabular}{c|cc}
\toprule
 & \multicolumn{2}{c}{New $Z'$} \\
\cmidrule(lr){2-3}
$M_{Z'}$ [GeV] & $g_*^{\rm EFT}$ & $g_*^{\rm EFT}/g_*^{\rm full}$ \\
\midrule
 400  & 0.015 & 2.3 \\
 500  & 0.018 & 1.4 \\
 700  & 0.025 & 1.2 \\
 1000 & 0.036 & 1.1 \\
 2000 & 0.073 & 1.0 \\
 5000 & 0.18  & 1.0 \\
\bottomrule
\end{tabular}
\caption{The coupling $g_*$ required for an optimistic $5\sigma$ discovery in $e^+e^-\to\mu^+\mu^-$ at $\sqrt{s}=365$~GeV with $3.12~\text{ab}^{-1}$ in the dimension-6 EFT, and its ratio to the result with the full $Z'$ propagator.
The leptophilic and $L_\mu - L_e$ models agree to $\lesssim 0.2\%$, since the only difference is the sign on the interference. 
}
\label{tab:eft_vs_full}
\end{table}

In the intermediate regime, where the mass of the new particle is too heavy for direct production, but not parametrically heavy to be fully captured by the EFT, modelling the interference and shape is the best probe of new dynamics. 

\subsection{Oblique parameters and the location of zeros}
Let us consider oblique $W$ and $Y$ parameters in the context of `Universal' UV completions of the SM.
These parameters encode the interference of new physics with the electroweak and hypercharged current, providing an obvious setting for the consideration of amplitude zeros.  As shown in \cite{McCullough:2023szd}, in the cases for which it is physical to parameterise new physics effects in this way these parameters will be strictly positive, hence interference is destructive.

\begin{figure}[t]
\begin{center}
\includegraphics[width=.47\textwidth]{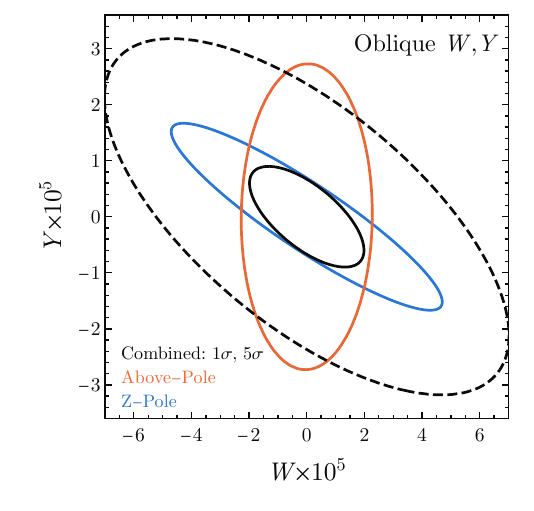} \includegraphics[width=.47\textwidth]{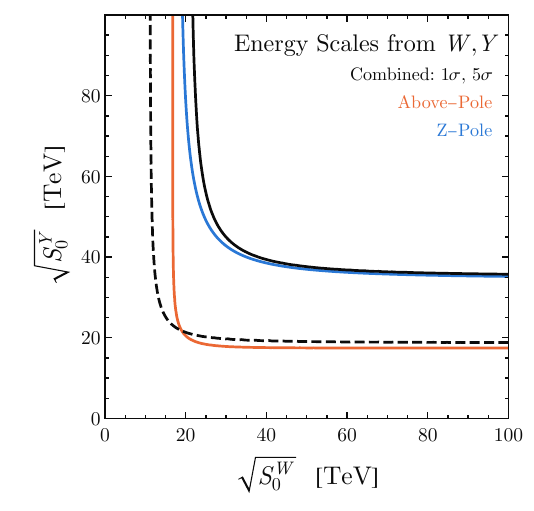} 
\caption{Left: Projected FCC-ee constraints on the $W$ and $Y$ parameters from Z-pole (blue) and $WW$, $ZH$ and $t \overline{t}$ runs (red).  The combined projections are also shown, with a $5\sigma$ contour to guide the eye.  Right:  The $W$ and $Y$ plane mapped into an upper limit on the energy scale of an associated amplitude zero or new states, under the assumption that $W$ and $Y$ are positive.} 
\label{fig:WY}
\end{center}
\end{figure}

In the left panel of Fig.~\ref{fig:WY} we show the projected exclusion sensitivity of FCC-ee measurements, reproduced from the analysis of \cite{Greljo:2024ytg}.  For the sake of comparison we have not restricted to positive values.

Suppose a non-zero hint for new physics emerged in the form of a non-zero extraction of $W$ or $Y$ at FCC-ee.  Given that present constraints are at the $\mathcal{O}(10^{-4})$ level such a hint could be as strong as a $5\sigma$ tension. For simplicity we consider the case of $W$.  In such a scenario, under the assumption of a universal UV-completion, we would map a measured value into the UV parameter combination
\begin{equation}
W = \frac{g_\star^2 M_W^2}{g^2 M^2} ~~.
\end{equation}
From this measurement alone a degeneracy between coupling and mass remains, making an unambiguous extraction of the mass of new physics impossible.  Upper bounds may be found from appealing to perturbative unitarity, leading to a limit such as $M \lesssim 4 \pi M_W / g \sqrt{w}$.  Alternatively, if the dimension-8 term could also be measured from the energy-distribution a stronger upper limit on the mass scale could be derived due to the convergence of the momentum expansion, as demonstrated in Ref.~\cite{Englert:2019zmt}.

Interestingly, measuring $W$ alone does not enable an unambiguous extraction of the relevant mass scale, however it does instead allow one to extract an upper limit on the energy scale at which the amplitude zero \emph{or} the new physics itself must arise.  For the former this is physically relevant as it marks the rate minimum.  Following Ref.~\cite{Englert:2019zmt,McCullough:2023szd}, under the assumption that the UV theory is universal and thus the UV density of states is positive and the amplitude zero lies below the new mass thresholds, we may parameterise the UV contribution to the IR propagator as
\begin{equation}
\delta \mathcal{A}(s) = \frac{1}{s-M_W^2} -\frac{W}{M_W^2} \sum_{n=0} c_n \left(\frac{s}{M^2} \right)^{n} ~~,
\end{equation}
where $c_0=1$, $c_n\geq 0$ and $c_n\geq c_{n+1}$ with the inequality saturated for tree-level exchange in the UV.  We thus find that even for non-perturbative universal UV-completions the amplitude zero must appear at or below the energy scale
\begin{equation}
s_0 \leq M_W^2 \left( \frac{1}{W} + 1 \right)  ~~.\label{eq:zeroornew}
\end{equation}
It is remarkable it is possible to bound the energy scale of an amplitude zero.  The only caveat to this argument is if there is no zero below the mass scale of new physics, however in that case the new physics scale is below the scale in Eq.~\ref{eq:zeroornew}, thus Eq.~\ref{eq:zeroornew} represents a `no-lose' scale for new physics in the event of a non-vanishing measurement of the W-parameter.\footnote{Note that Eq.~\ref{eq:zeroornew} demonstrates that a measurement of $W \sim 2\times 10^{-5}$ would typically place the upper scale for the zero well beyond the kinematic reach of FCC-ee.}

In the right hand panel of Fig.~\ref{fig:WY} the location of the amplitude zero bound corresponding to the left-hand panel is shown.  Interestingly, any deviation corresponding to a $5\sigma$ deviation at FCC-ee would point towards the existence of an amplitude zero (or new states) at $\sqrt{s_0} \lesssim 20$ TeV, within the kinematic range of FCC-hh.

\subsection{Zeros in kinematic reach}
\label{sec:zeroskinematic}
We now move to the scenario involving realistic high-energy facilities.  As argued in Sec.~\ref{sec:low_eng}, the shape around the zero is controlled by the low energy observables that specify the energy of the dip, its depth, and the curvature. 
In the following we perform a phenomenological study of a case in which the features of the zero are within kinematical reach. 
For illustrative purposes, we consider a muon collider running at $\sqrt{s}=10\,\text{ TeV}$. 
Due to initial state radiation, the collider provides partonic collisions at $\sqrt{\hat s}$ lower than the collider energy, providing a luminosity that is captured by leptonic parton distribution functions \cite{Garosi:2023bvq}. 
We will assume that the amplitude for $\mu^+\mu^-\to e^+e^-$ has a zero $s_0$ close to this range of energies, and try to understand how to characterize its phenomenology and interpretation.\footnote{Here the \textit{zero} denotes the real-energy dip in the total rate associated with the complex amplitude zero due to the heavy resonance width, but otherwise we do not include a complete calculation of virtual and radiative corrections.} 

The SM amplitude is dominated by photon and $Z$ exchanges, so at high energies the SM coupling among the different chirality configurations is $g^{2,SM}_{\chi\chi^\prime}=e^2+g_\chi^\mu g_{\chi^\prime}^e$, as discussed in Sec.~\ref{sec:chiralcouplingsgeneral}. We assume the heavy resonance to be exactly aligned to the SM, thus we assume the residue of the UV pole $g^{2,BSM}_{\chi\chi^\prime}$ to be $g^{2,BSM}_{\chi\chi^\prime}=\frac{g_\star^2}{g^2}g^{2,SM}_{\chi\chi^\prime}$. Thus while the template used implies a model with this alignment, the similarity of diagonal SM coefficients allows for a single vector resonance to produce a near zero in the same chirality channel, as can be seen in Fig.~\ref{fig:chiralzeros}, even if the coupling matrix $g^{2,BSM}_{\chi\chi^\prime}$ for a single resonance exchange is rank one. 

Using the leptonic PDFs of \cite{Garosi:2023bvq}, we compute the $\mu^+\mu^-\to e^+e^-$ cross section as a function of the partonic center of mass energies $\sqrt{\hat s}$, which coincide with the invariant mass of the electron pair. We bin $\sqrt{\hat s}$ from 5~TeV to 10~TeV in bins of 200~GeV and measure the observed cross section $\sigma$ normalized to the SM rate $\sigma_{SM}$. For 10$~\text{ab}^{-1}$ of luminosity, from the expected events in each bin as predicted in the SM we obtain a statistical uncertainty of order $15\%$ in the first bins, which decreases as the partonic center-of-mass energy increases until $\sim 5\%$ in the next-to-last bin and $\sim 1\%$ at the last bin. We assume a $2\%$ systematic uncertainty, which only significantly affects the last bin.

\begin{figure}[t]
    \centering
    \includegraphics[width=1\linewidth]{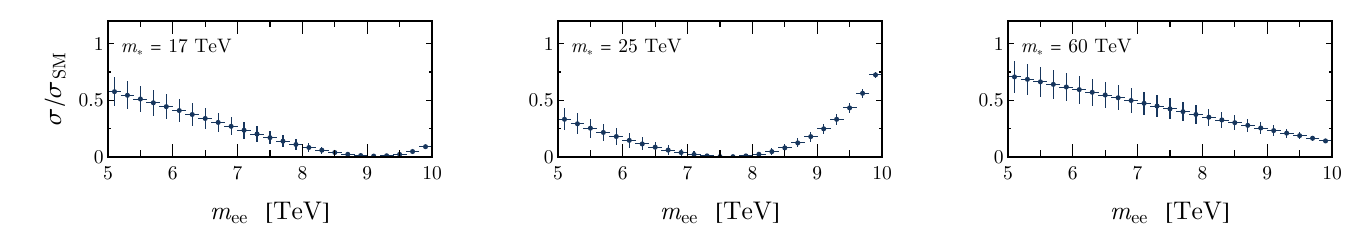}
    \includegraphics[width=1\linewidth]{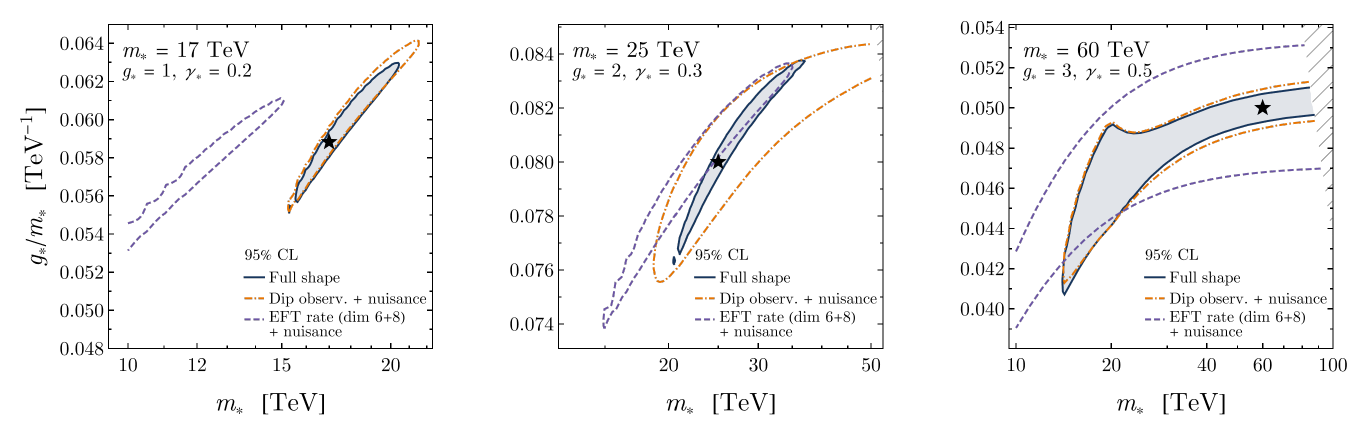}
    \caption{The three panels correspond to the single-resonance models with the parameters specified in the insets. Top: Cross section normalized to the SM rate. Bottom: Interpretation of the low energy fit in terms of $(m_\star,g_\star)$ under three different assumptions, see text.}
    \label{fig:muoncollfit}
\end{figure}

We assume the amplitude to be given by the single-resonance model of Eq.~\ref{eq:amplitudedeepwidth}, with couplings aligned in chirality with the SM. 
We inject three different signal points, corresponding to $(m_\star,g_\star,\gamma_\star)$ given by $(17\,\text{TeV}, 1, 0.2)$, $(25\,\text{TeV}, 2, 0.3)$ and $(60\,\text{TeV}, 3, 0.5)$. The first two points lead to a zero at energies $\sqrt{s_0}\sim 9\,\text{TeV}$ and $\sqrt{s_0}\sim 7.5\,\text{TeV}$, respectively, as can be seen in the top panels of Fig.~\ref{fig:muoncollfit}. The last point leads to a zero at $\sqrt{s_0}\sim 12\,\text{TeV}$, which is outside the kinematic range. 
All three benchmarks give clear departures from the SM rate. For the first two, the fit determines both $s_0$ and $\mathcal K$ precisely. For the third, it primarily constrains $g_*/m_*$.
We compare two fits to the dip with a truncated EFT fit.

First, we assume the functional form of $\sigma/\sigma_\text{SM}=|A(s)|^2$ as given by the single resonance model in Eq.~\ref{eq:amplitudedeepwidth}. In this particular case, it coincides with the underlying functional form, and thus determines the optimal interpretation of the IR measurements. Using the map in Eq.~\ref{eq:irtouvmappingsingleresonance} (we use the form to all orders in $\gamma_\star^2$), we obtain the grey area of the lower panels in Fig.~\ref{fig:muoncollfit}. 
We show the $95\%$ confidence regions defined as $\Delta\chi^2=5.99$ relative to the best fit point, using a binned Poisson likelihood with independent 2\% uncertainties in each bin and profiling over the remaining parameters. 
The best fit point coincides with the truth value, and in the first two cases, where both the zero and the curvature are measured with precision, the mass and coupling are determined with $\sim$ few \% precision. 
In the third panel, only the combination $g_\star/m_\star$ is determined with precision, with the mass mostly undetermined. That it is only this combination can be understood either by noting that it is the combination that enters at dimension 6 level, or by noting that it is the combination that controls the zero $s_0$, with the curvature being mostly undetermined.  The hatched region denotes the $\Gamma/m_\star=1$ bound as computed from only the $\mu$ and $e$ channels. 

Second, we measure the dip observables $(s_0,\mathcal K, \gamma)$, which determine the amplitude squared around the zero, $|A(s)|^2\simeq \gamma^2 + \mathcal K^2 \frac{(s-s_0)^2}{s_0^2}+\dots$. We do so without assuming the functional form in Eq.~\ref{eq:amplitudedeepwidth}. The reason is that this form is only true for the specific case of a single resonance, and generally not true for the more general case of the positive spectral density of Sec.~\ref{sec:strongcoupling}. What we do is to perform a fit on these three parameters, and profile over a cubic term $\theta (s-s_0)^3$, as a purely quadratic form does not fit the observed shapes. We interpret the obtained $(s_0,\mathcal K)$ in terms of $(m_\star^2,g_\star^2)$ as in the single resonance map in Eq.~\ref{eq:irtouvmappingsingleresonance}, which leads to the orange contours in Fig.~\ref{fig:muoncollfit}. 

We compare this with a truncated EFT fit, where we write $|A(s)|^2\simeq 1-2c_6 s+ (c_6^2-2c_8)s^2+(2c_6 c_8-2c_{10})s^3+\dots$, and profile over $c_{10}$. We consider $c_{10}$ only subject to the positivity of the total rate, not positivity or monotonicity of $c_{10}$ itself. In this way, we fit a cubic form keeping two EFT parameters, $c_6$ and $c_8$, in a consistent manner in the total rate. Then, we extract the UV parameters by interpreting $m_\star^2=c_6/c_8$ and $g_\star^2/g^2=c_6^2/c_8$. We obtain the dashed purple line in Fig.~\ref{fig:muoncollfit}. 
We observe that in the last panel this gives results compatible with the other procedure for $g_\star/m_\star$. The reason is that the dimension-8 operator is further suppressed by ${\hat s}/m_\star^2$ relative to the dimension-6 one, which in this case is close to the limit of precision. In the first panel, the opposite is true, and the truncated EFT expansion does not provide a satisfactory inference of the UV parameters. This is to be contrasted with the orange curve of the fit on the zero, which correctly described the benchmark point. 
The second panel is a situation where the truncated EFT description is good enough to give a good approximation to the true $(m_\star,g_\star)$ point.

\section{Conclusions}
\label{sec:conclusions}
A central question for the precision era is how heavy new physics manifests at energies below its mass.
We demonstrate the importance of amplitude zeros in this context:  Destructive interference between light and heavy dynamics can cancel the amplitude at energies parametrically below the heavy threshold.
In the single-resonance model, whenever the heavy contribution has the same sign residue as the SM amplitude the zero is guaranteed to lie below the new mass threshold, and for a coupling hierarchy $\gss/g^2 \gg 1$ it sits at $s_0 \simeq m_\star^2\, g^2/\gss$, without requiring the new dynamics to be non-perturbative.
In realistic settings the exact zero is lifted by absorptive effects and diluted by the sum over helicity channels, but the interference survives as a dip, or more generally as a smooth deficit or excess in measured distributions, set by the sign of the UV residue.

Amplitude zeros are compatible with an EFT description, but they sharpen the question of where it can be truncated.
The leading dimension-6 term already predicts the location of a parametrically light zero. 
Near the cancellation, the squared dimension-6 amplitude must be retained in the rate. 
The SM interference with the dimension-8 amplitude enters at the same EFT order, although the coupling hierarchy suppresses it in our example.
Additionally, absorptive effects enter only through loops of the EFT interactions.
At the same time, the shape of an accessible dip turns this into a measurable question:  Its position and curvature determine the relative size of the EFT terms.
For certain UV completions with a positive spectral density, the zero position $s_0$ and curvature $\mathcal K$ bound the heavy threshold, $m_{\rm th}^2 \le s_0\,\mathcal K/(\mathcal K-1)$, independently of the coupling and of the convergence of the EFT expansion.

This picture suggests a path for heavy new physics through successive generations of collider experiments.
Zeros are already relevant for searches at the HL-LHC.
For a sequential $W'$, the zero reshapes the transverse-mass distribution well below the resonance, moving sensitivity away from the highest bins.
Analyses that add signal and background incoherently can therefore misjudge both the signal shape and the regions that drive sensitivity.
At the FCC-ee, the reach of $e^+e^-\to\mu^+\mu^-$ to leptophilic mediators extends far beyond LEP-II once interference is included.
This holds for deficits and excesses alike, and is particularly important in the region of parameter space poorly modeled by low-order EFT.
FCC-ee can also indicate where to look next.
Under universal-theory assumptions, a significant non-zero discovery of $W$ or $Y$ would place an amplitude zero or the new states themselves below roughly $20$~TeV, within reach of FCC-hh or a high-energy muon collider.
Such a machine could directly access the dip, and the position, depth and curvature could then potentially determine the mass and coupling of the heavy state.

Several steps remain before this sensitivity can be realized in a full search.
The most immediate is a realistic treatment of uncertainties.
Our FCC-ee reach depends strongly on how well the absolute rate is known, and the $W'$ example shows how correlated systematics can change which bins dominate.
Existing LHC and LEP measurements offer a natural testing ground, and recasting them with coherent SM and BSM predictions would serve as a proof of concept.

On the theory side, there are also improvements to be made for a more general analysis strategy.
Precise SM predictions are essential, since radiative corrections can produce deficits that mimic destructive interference \cite{Chiesa:2013yma,Alioli:2016fum,Denner:2019vbn}.
They also determine how far absorptive effects lift a zero into a dip, which sets how deep a feature one can hope to observe.
Additionally, our examples assumed $s$-channel exchange of a single current.
Extending them to processes with $t$-channel contributions or to several interfering currents would clarify how generic a parametrically light zero is, as well as open new final states for complementary analyses.

As colliders move into an era defined by precision rather than energy, heavy new physics may first appear through its interference with the SM.
The absence of a resonance, or even of a resolved dip, does not mean the effects of heavy dynamics are small.
An amplitude is shaped by its zeros as well as its poles, and while an exact zero is rarely observable, its imprint can appear at energies well below the new physics pole.
Searching for these interference effects offers a complementary route to discovery and enhances the precision program of the coming decades in its role as a probe of dynamics well beyond kinematic reach.

\acknowledgments{The authors are grateful to Bego\~na de la Cruz for informative discussions and comments on an early draft.  ChatGPT was used for verification of final versions of the preprint draft in the form of a putative referee.}

\appendix

\section{Width effects in the EFT description}

\label{sec:width}

There has been considerable discussion of how higher-order contributions should be treated in EFT analyses of collider data, for instance whether squared dimension-6 or dimension-8 terms must be retained.\footnote{See e.g.\ Ref.~\cite{Contino:2016jqw} for the relevant power counting.}
Amplitude zeros raise a different issue.
At a tree-level zero the real part of the amplitude vanishes, so the rate is controlled by its absorptive part. 
Note that no order of the tree-level derivative expansion can generate the absorptive effects.  In the full theory the effect is simply related to the width of the heavy mediator.
Here we show that in the EFT the same physics arises from a class of IR loop diagrams, reproducing the UV width effect without direct reference to the UV, as an EFT ought to.
For narrow resonances these contributions are negligible almost everywhere, but near the zero they determine how far the amplitude is lifted.

\subsection{A toy model}

Consider a more involved toy model to illustrate how this resolution relates the width of the mediator to the EFT loops. 
A pair of massless real scalar fields $\phi_{1,2}$, each with a $\mathcal{Z}_2$ symmetry, couple to a heavy scalar field $\Phi_{H}$ with trilinear coupling $A_H$:
\begin{equation}
\mathcal{L}_{\rm int} =     - \frac{A_H}{2}\left(\phi_1^2 + \phi_2^2\right)\Phi_H\,.
\end{equation}
This model isn't natural, but that is not important for illustrating the overall lesson.

The amplitude contribution from $\Phi_{H}$ for $\phi_1 \phi_1 \to \phi_2 \phi_2$ scattering is
\begin{eqnarray}
\mathcal{M}(s) & = &   \frac{A_H^2}{s-M_H^2+ i \text{Im} \left[ \Sigma_H (s) \right]} \nonumber \\
& \approx &  \frac{A_H^2}{s-M_H^2+ i M_H \Gamma_H(s)}~~,
\end{eqnarray}
where in the last line in the vicinity of the resonance the (one-loop) self-energy is related to the tree-level width via the optical theorem as
\begin{equation}
\text{Im} \left[ \Sigma_H (M_H^2) \right] =   M_H \Gamma_H (M_H^2) ~~.
\end{equation}

From the EFT perspective, one ought to be able to capture all UV effects from $\Phi_H$ in the IR with the EFT.
At tree-level, the UV matching clearly generates the infinite tower of effective operators arising from the $\Box/M_H^2$ expansion of
\begin{equation}
\mathcal{L}_{\text{EFT}} \supset \frac{A_H^2}{8} (\phi_1^2+\phi_2^2) \frac{1}{\Box+M_H^2} (\phi_1^2+\phi_2^2)~~.
\end{equation}

In the IR this tower of operators generates the tree-level scattering amplitude, however, the width effects are absent. 

We now show how the UV width effect maps within the EFT to IR loop effects which give rise to the same absorptive component of the EFT amplitude.

\begin{figure}[t]
\begin{center}
\includegraphics[width=.9\textwidth]{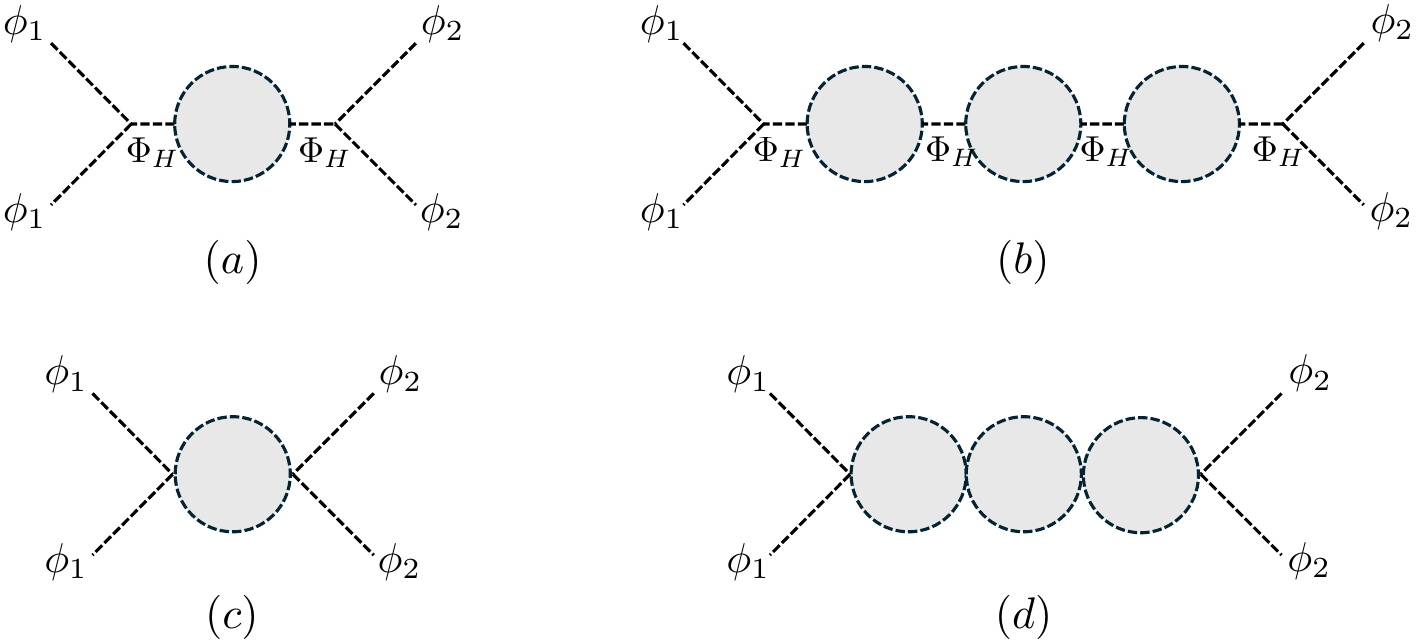} 
\caption{\label{fig:UVloop} UV width effects, wherein the gray bubble includes all states to which $\Phi_H$ couples. 
The top panels show the full UV calculation, and the bottom panels the equivalent EFT contribution.
The optical theorem relates the imaginary part of the loop to the $\Phi_H$ width and the right hand panels show an example of the classes of diagrams contributing to the resummation.} 
\end{center}
\end{figure}

Through the optical theorem in the UV, we may attribute the width effects to the resummation of a class of diagrams shown in Fig.~\ref{fig:UVloop} (a), wherein all possible decay products of $\Phi_H$ are included in the bubble, namely $\phi_{1,2}$.
This width is
\begin{equation}
\Gamma_H = \frac{A_H^2}{16 \pi M_H} + \text{other modes}~~.
\end{equation}
This gives the leading contribution to the amplitude proportional to the width, however the full effect in the denominator of the propagator arises after resummation of all diagrams with an internal loop in the propagator as shown in (b).  For the case in which matching is perturbatively calculable and the UV resonance is sufficiently narrow, only the leading-order width correction is necessary.  In this case at lowest relevant orders in the momentum expansion of the $\phi_1 \phi_1 \to \phi_2 \phi_2$ UV amplitude is given by
\begin{equation}
\mathcal{M}(s)  =   - \frac{A_H^2}{M_H^4} \left(M_H^2+ i \text{Im} \left[ \Sigma_{\text{EFT}} (s) \right] \right) ~~,
\end{equation}
where
\begin{equation}
 \text{Im} \left[ \Sigma_{\text{EFT}} (s) \right] = \text{Im} \left[ \Sigma_H (s) \right] ~~
\end{equation}
where an interpretation in terms of UV particle widths is possible but not necessary within the EFT.

By comparing with the UV diagrams we now see how the effects of the UV particle width arise in the EFT, since the same classes of diagram appear, however with $\Phi_H$ integrated out as in Fig.~\ref{fig:UVloop} (c).  It is the resummation of these diagrams which captures the heavy particle width effects (d) in the IR EFT.  This effect can be calculated within the low-energy EFT such that, given a local EFT operator description, the one-loop effects which may become dominant in regions of tree-level amplitude zeros may be reliably calculated without any knowledge of the UV.  Within the EFT alone one would not na\"ively consider this class of diagrams to be related to a UV particle width, however the latter interpretation reveals their crucial importance in correctly accounting for the complex structure of amplitudes in the IR, particularly relevant in regions of interference. 

Integrating out the heavy field leads to a quartic $\mathcal{L}\supset \frac{A_H^2}{4 M_H^2}\phi_1^2\phi_2^2$. The resummation of the diagrams in Fig.~\ref{fig:UVloop} (d), given by $-\frac{A_H^2}{M_H^2}\left( 1+ i \frac{A_H^2/M_H^2}{16\pi} +\dots \right)$ leads to a term in the amplitude of the form
\begin{equation}
-\frac{A_H^2}{M_H^2} \frac{1}{1-i \frac{A_H^2/M_H^2}{16\pi}}\, .
\end{equation}
Thus the width is fully captured within the EFT by the resummation of a particular class of diagrams. 
Notice that by consistency, not only the $\phi_1^2\phi_2^2$ operator is related to the width effect, but also $\phi_1^4$ and $\phi_2^4$, from the same resummation. 

In general models it is important to retain any energy dependence in $\text{Im} \left[ \Sigma_H (s) \right]$.  Thus the strict interpretation is between the relevant EFT one-loop diagrams and the general off-shell width of the UV particle, since the internal propagators in Figs.~\ref{fig:UVloop} (a) and (b) are not on-shell.

For our simple example this energy-dependence is trivial, however we show this may not be the case through inspecting some simple additional examples, employing the calculations of \cite{Banks:2020gpu}.  For example if the UV states generate a coupling of the light fields $\phi_{1,2}$ to some operator $\mathcal{O}'_L$ as $\frac{A_H}{2} (\phi_1^2+\phi_2^2) \mathcal{O}'_L$, where $\mathcal{O}'_L$ is
\begin{equation}
(A)~~~~ \frac{1}{2} \phi^2 ~~,~~(B)~~~~ \frac{1}{2} \psi^2  ~~,~~(C)~~~~ \partial_\mu \phi \partial^\mu \phi ~~,~~(D)~~~~ \frac{1}{n!} \phi^n
\end{equation}
then the effective contribution of the operator $\mathcal{O}'_L$ to the self-energy of the operator $(\phi_1^2+\phi_2^2)$ is
\begin{equation}
\text{Im} \left[ \Sigma_H (s) \right] = \frac{A_H^2}{32 \pi} \left\{ ~~ 1 ~~,~~ 2 s ~~,~~ s^2 ~~,~~  \frac{2 s^{n-2}}{(4\pi)^{2(n-2)} n! (n-1)! (n-2)!} ~~ \right\} ~~,
\end{equation}
with energy-growing effects for marginal and irrelevant interactions.

\bibliographystyle{JHEP}
\bibliography{refs.bib}

\providecommand{\href}[2]{#2}\begingroup\raggedright\begin{thebibliography}{10}

\bibitem{ZurbanoFernandez:2020cco}
I.~Zurbano~Fernandez et~al., \emph{{High-Luminosity Large Hadron Collider
  (HL-LHC): Technical design report}}, .

\bibitem{FCC:2018byv}
{\scshape FCC} collaboration, A.~Abada et~al., \emph{{FCC Physics
  Opportunities}: {Future Circular Collider Conceptual Design Report Volume
  1}}, \href{https://doi.org/10.1140/epjc/s10052-019-6904-3}{\emph{Eur. Phys.
  J. C} {\bfseries 79} (2019) 474}.

\bibitem{FCC:2018evy}
{\scshape FCC} collaboration, A.~Abada et~al., \emph{{FCC-ee: The Lepton
  Collider}: {Future Circular Collider Conceptual Design Report Volume 2}},
  \href{https://doi.org/10.1140/epjst/e2019-900045-4}{\emph{Eur. Phys. J. ST}
  {\bfseries 228} (2019) 261--623}.

\bibitem{FCC:2025lpp}
{\scshape FCC} collaboration, M.~Benedikt et~al., \emph{{Future Circular
  Collider Feasibility Study Report: Volume 1, Physics, Experiments,
  Detectors}},
  \href{https://doi.org/10.1140/epjc/s10052-025-15077-x}{\emph{Eur. Phys. J. C}
  {\bfseries 85} (2025) 1468},
  [\href{https://arxiv.org/abs/2505.00272}{{\ttfamily 2505.00272}}].

\bibitem{Contino:2016jqw}
R.~Contino, A.~Falkowski, F.~Goertz, C.~Grojean and F.~Riva, \emph{{On the
  Validity of the Effective Field Theory Approach to SM Precision Tests}},
  \href{https://doi.org/10.1007/JHEP07(2016)144}{\emph{JHEP} {\bfseries 07}
  (2016) 144}, [\href{https://arxiv.org/abs/1604.06444}{{\ttfamily
  1604.06444}}].

\bibitem{Gaemers:1984sj}
K.~J.~F. Gaemers and F.~Hoogeveen, \emph{{Higgs Production and Decay Into Heavy
  Flavors With the Gluon Fusion Mechanism}},
  \href{https://doi.org/10.1016/0370-2693(84)91711-8}{\emph{Phys. Lett. B}
  {\bfseries 146} (1984) 347--349}.

\bibitem{Bella:2010sc}
G.~Bella, E.~Etzion, N.~Hod, Y.~Oz, Y.~Silver and M.~Sutton, \emph{{A Search
  for heavy Kaluza-Klein electroweak gauge bosons at the LHC}},
  \href{https://doi.org/10.1007/JHEP09(2010)025}{\emph{JHEP} {\bfseries 09}
  (2010) 025}, [\href{https://arxiv.org/abs/1004.2432}{{\ttfamily 1004.2432}}].

\bibitem{Bai:2014fkl}
Y.~Bai and W.-Y. Keung, \emph{{Can vanishing mass-on-shell interactions
  generate a dip at colliders?}},
  \href{https://doi.org/10.1142/S0217751X15501201}{\emph{Int. J. Mod. Phys. A}
  {\bfseries 30} (2015) 1550120},
  [\href{https://arxiv.org/abs/1407.6355}{{\ttfamily 1407.6355}}].

\bibitem{Fiaschi:2021sin}
J.~Fiaschi, F.~Giuli, F.~Hautmann and S.~Moretti, \emph{{Enhancing the Large
  Hadron Collider sensitivity to charged and neutral broad resonances of new
  gauge sectors}}, \href{https://doi.org/10.1007/JHEP02(2022)179}{\emph{JHEP}
  {\bfseries 02} (2022) 179},
  [\href{https://arxiv.org/abs/2111.09698}{{\ttfamily 2111.09698}}].

\bibitem{BaronMoreno:2026pjr}
D.~A. Baron~Moreno, C.~Englert and Y.~Peters, \emph{{Big Dipper, Help Me Find A
  Way -- Dip-hunting at hadron colliders}},
  \href{https://arxiv.org/abs/2604.25407}{{\ttfamily 2604.25407}}.

\bibitem{Frixione:2020hqz}
S.~Frixione, L.~Roos, E.~Ting, E.~Vryonidou, M.~White and A.~G. Williams,
  \emph{{Model-independent approach for incorporating interference effects in
  collider searches for new resonances}},
  \href{https://doi.org/10.1140/epjc/s10052-020-08763-5}{\emph{Eur. Phys. J. C}
  {\bfseries 80} (2020) 1174},
  [\href{https://arxiv.org/abs/2008.07751}{{\ttfamily 2008.07751}}].

\bibitem{Barcelo:2010bm}
R.~Barcelo and M.~Masip, \emph{{Extra Higgs bosons in $t\bar{t}$ production at
  the LHC}}, \href{https://doi.org/10.1103/PhysRevD.81.075019}{\emph{Phys. Rev.
  D} {\bfseries 81} (2010) 075019},
  [\href{https://arxiv.org/abs/1001.5456}{{\ttfamily 1001.5456}}].

\bibitem{Bernreuther:2015fts}
W.~Bernreuther, P.~Galler, C.~Mellein, Z.~G. Si and P.~Uwer, \emph{{Production
  of heavy Higgs bosons and decay into top quarks at the LHC}},
  \href{https://doi.org/10.1103/PhysRevD.93.034032}{\emph{Phys. Rev. D}
  {\bfseries 93} (2016) 034032},
  [\href{https://arxiv.org/abs/1511.05584}{{\ttfamily 1511.05584}}].

\bibitem{Hespel:2016qaf}
B.~Hespel, F.~Maltoni and E.~Vryonidou, \emph{{Signal background interference
  effects in heavy scalar production and decay to a top-anti-top pair}},
  \href{https://doi.org/10.1007/JHEP10(2016)016}{\emph{JHEP} {\bfseries 10}
  (2016) 016}, [\href{https://arxiv.org/abs/1606.04149}{{\ttfamily
  1606.04149}}].

\bibitem{Carena:2016npr}
M.~Carena and Z.~Liu, \emph{{Challenges and opportunities for heavy scalar
  searches in the $ t\overline{t} $ channel at the LHC}},
  \href{https://doi.org/10.1007/JHEP11(2016)159}{\emph{JHEP} {\bfseries 11}
  (2016) 159}, [\href{https://arxiv.org/abs/1608.07282}{{\ttfamily
  1608.07282}}].

\bibitem{Djouadi:2016ack}
A.~Djouadi, J.~Ellis and J.~Quevillon, \emph{{Interference effects in the
  decays of spin-zero resonances into $\gamma \gamma$ and $ t\overline{t} $}},
  \href{https://doi.org/10.1007/JHEP07(2016)105}{\emph{JHEP} {\bfseries 07}
  (2016) 105}, [\href{https://arxiv.org/abs/1605.00542}{{\ttfamily
  1605.00542}}].

\bibitem{Maina:2015ela}
E.~Maina, \emph{{Interference effects in Heavy Higgs production via gluon
  fusion in the Singlet Extension of the Standard Model}},
  \href{https://doi.org/10.1007/JHEP06(2015)004}{\emph{JHEP} {\bfseries 06}
  (2015) 004}, [\href{https://arxiv.org/abs/1501.02139}{{\ttfamily
  1501.02139}}].

\bibitem{Kauer:2015hia}
N.~Kauer and C.~O'Brien, \emph{{Heavy Higgs signal{\textendash}background
  interference in $gg\rightarrow VV$ in the Standard Model plus real singlet}},
  \href{https://doi.org/10.1140/epjc/s10052-015-3586-3}{\emph{Eur. Phys. J. C}
  {\bfseries 75} (2015) 374},
  [\href{https://arxiv.org/abs/1502.04113}{{\ttfamily 1502.04113}}].

\bibitem{Ballestrero:2015jca}
A.~Ballestrero and E.~Maina, \emph{{Interference Effects in Higgs production
  through Vector Boson Fusion in the Standard Model and its Singlet
  Extension}}, \href{https://doi.org/10.1007/JHEP01(2016)045}{\emph{JHEP}
  {\bfseries 01} (2016) 045},
  [\href{https://arxiv.org/abs/1506.02257}{{\ttfamily 1506.02257}}].

\bibitem{Greiner:2016awe}
N.~Greiner, S.~H{\"o}che, G.~Luisoni, M.~Sch{\"o}nherr and J.-C. Winter,
  \emph{{Full mass dependence in Higgs boson production in association with
  jets at the LHC and FCC}},
  \href{https://doi.org/10.1007/JHEP01(2017)091}{\emph{JHEP} {\bfseries 01}
  (2017) 091}, [\href{https://arxiv.org/abs/1608.01195}{{\ttfamily
  1608.01195}}].

\bibitem{Dawson:2015haa}
S.~Dawson and I.~M. Lewis, \emph{{NLO corrections to double Higgs boson
  production in the Higgs singlet model}},
  \href{https://doi.org/10.1103/PhysRevD.92.094023}{\emph{Phys. Rev. D}
  {\bfseries 92} (2015) 094023},
  [\href{https://arxiv.org/abs/1508.05397}{{\ttfamily 1508.05397}}].

\bibitem{Jung:2015etr}
S.~Jung, J.~Song and Y.~W. Yoon, \emph{{How Resonance-Continuum Interference
  Changes 750 GeV Diphoton Excess: Signal Enhancement and Peak Shift}},
  \href{https://doi.org/10.1007/JHEP05(2016)009}{\emph{JHEP} {\bfseries 05}
  (2016) 009}, [\href{https://arxiv.org/abs/1601.00006}{{\ttfamily
  1601.00006}}].

\bibitem{Craig:2016ygr}
N.~Craig, J.~Hajer, Y.-Y. Li, T.~Liu and H.~Zhang, \emph{{Heavy Higgs bosons at
  low $\tan \beta$: from the LHC to 100 TeV}},
  \href{https://doi.org/10.1007/JHEP01(2017)018}{\emph{JHEP} {\bfseries 01}
  (2017) 018}, [\href{https://arxiv.org/abs/1605.08744}{{\ttfamily
  1605.08744}}].

\bibitem{Martin:2016bgw}
S.~P. Martin, \emph{{Signal-background interference for a singlet spin-0
  digluon resonance at the LHC}},
  \href{https://doi.org/10.1103/PhysRevD.94.035003}{\emph{Phys. Rev. D}
  {\bfseries 94} (2016) 035003},
  [\href{https://arxiv.org/abs/1606.03026}{{\ttfamily 1606.03026}}].

\bibitem{Bhattiprolu:2020zoq}
P.~N. Bhattiprolu and S.~P. Martin, \emph{{Signal-background interference for
  digluon resonances at the Large Hadron Collider}},
  \href{https://doi.org/10.1103/PhysRevD.102.015016}{\emph{Phys. Rev. D}
  {\bfseries 102} (2020) 015016},
  [\href{https://arxiv.org/abs/2004.06181}{{\ttfamily 2004.06181}}].

\bibitem{Rizzo:2009pu}
T.~G. Rizzo, \emph{{Indirect Searches for Z-prime-like Resonances at the LHC}},
  \href{https://doi.org/10.1088/1126-6708/2009/08/082}{\emph{JHEP} {\bfseries
  08} (2009) 082}, [\href{https://arxiv.org/abs/0904.2534}{{\ttfamily
  0904.2534}}].

\bibitem{Accomando:2013sfa}
E.~Accomando, D.~Becciolini, A.~Belyaev, S.~Moretti and
  C.~Shepherd-Themistocleous, \emph{{Z' at the LHC: Interference and Finite
  Width Effects in Drell-Yan}},
  \href{https://doi.org/10.1007/JHEP10(2013)153}{\emph{JHEP} {\bfseries 10}
  (2013) 153}, [\href{https://arxiv.org/abs/1304.6700}{{\ttfamily 1304.6700}}].

\bibitem{Accomando:2016mvz}
E.~Accomando, D.~Barducci, S.~De~Curtis, J.~Fiaschi, S.~Moretti and C.~H.
  Shepherd-Themistocleous, \emph{{Drell-Yan production of multi Z'-bosons at
  the LHC within Non-Universal ED and 4D Composite Higgs Models}},
  \href{https://doi.org/10.1007/JHEP07(2016)068}{\emph{JHEP} {\bfseries 07}
  (2016) 068}, [\href{https://arxiv.org/abs/1602.05438}{{\ttfamily
  1602.05438}}].

\bibitem{Baur:2007ck}
U.~Baur and L.~H. Orr, \emph{{High $p_{T}$ Top Quarks at the Large Hadron
  Collider}}, \href{https://doi.org/10.1103/PhysRevD.76.094012}{\emph{Phys.
  Rev. D} {\bfseries 76} (2007) 094012},
  [\href{https://arxiv.org/abs/0707.2066}{{\ttfamily 0707.2066}}].

\bibitem{Ferrario:2008wm}
P.~Ferrario and G.~Rodrigo, \emph{{Massive color-octet bosons and the charge
  asymmetries of top quarks at hadron colliders}},
  \href{https://doi.org/10.1103/PhysRevD.78.094018}{\emph{Phys. Rev. D}
  {\bfseries 78} (2008) 094018},
  [\href{https://arxiv.org/abs/0809.3354}{{\ttfamily 0809.3354}}].

\bibitem{Barcelo:2011vk}
R.~Barcelo, A.~Carmona, M.~Masip and J.~Santiago, \emph{{Stealth gluons at
  hadron colliders}},
  \href{https://doi.org/10.1016/j.physletb.2011.12.002}{\emph{Phys. Lett. B}
  {\bfseries 707} (2012) 88--91},
  [\href{https://arxiv.org/abs/1106.4054}{{\ttfamily 1106.4054}}].

\bibitem{Choudhury:2011cg}
D.~Choudhury, R.~M. Godbole and P.~Saha, \emph{{Dijet resonances, widths and
  all that}}, \href{https://doi.org/10.1007/JHEP01(2012)155}{\emph{JHEP}
  {\bfseries 01} (2012) 155},
  [\href{https://arxiv.org/abs/1111.1054}{{\ttfamily 1111.1054}}].

\bibitem{Dasgupta:2023zrh}
A.~Dasgupta, P.~S.~B. Dev, T.~Han, R.~Padhan, S.~Wang and K.~Xie,
  \emph{{Searching for heavy leptophilic Z': from lepton colliders to
  gravitational waves}},
  \href{https://doi.org/10.1007/JHEP12(2023)011}{\emph{JHEP} {\bfseries 12}
  (2023) 011}, [\href{https://arxiv.org/abs/2308.12804}{{\ttfamily
  2308.12804}}].

\bibitem{Mathews:2004xp}
P.~Mathews, V.~Ravindran, K.~Sridhar and W.~L. van Neerven,
  \emph{{Next-to-leading order QCD corrections to the Drell-Yan cross section
  in models of TeV-scale gravity}},
  \href{https://doi.org/10.1016/j.nuclphysb.2005.01.051}{\emph{Nucl. Phys. B}
  {\bfseries 713} (2005) 333--377},
  [\href{https://arxiv.org/abs/hep-ph/0411018}{{\ttfamily hep-ph/0411018}}].

\bibitem{Kumar:2009nn}
M.~C. Kumar, P.~Mathews, V.~Ravindran and A.~Tripathi, \emph{{Direct photon
  pair production at the LHC to order $\alpha_s$ in TeV scale gravity models}},
  \href{https://doi.org/10.1016/j.nuclphysb.2009.03.022}{\emph{Nucl. Phys. B}
  {\bfseries 818} (2009) 28--51},
  [\href{https://arxiv.org/abs/0902.4894}{{\ttfamily 0902.4894}}].

\bibitem{Calisto:2026pvv}
F.~Calisto, C.~Cheung, G.~N. Remmen, F.~Sciotti and M.~Tarquini, \emph{{On the
  Inverse Problem in Effective Field Theory}},
  \href{https://arxiv.org/abs/2604.15423}{{\ttfamily 2604.15423}}.

\bibitem{Berdine:2007uv}
D.~Berdine, N.~Kauer and D.~Rainwater, \emph{{Breakdown of the Narrow Width
  Approximation for New Physics}},
  \href{https://doi.org/10.1103/PhysRevLett.99.111601}{\emph{Phys. Rev. Lett.}
  {\bfseries 99} (2007) 111601},
  [\href{https://arxiv.org/abs/hep-ph/0703058}{{\ttfamily hep-ph/0703058}}].

\bibitem{Chiesa:2013yma}
M.~Chiesa, G.~Montagna, L.~Barz{\`e}, M.~Moretti, O.~Nicrosini, F.~Piccinini
  et~al., \emph{{Electroweak Sudakov Corrections to New Physics Searches at the
  LHC}}, \href{https://doi.org/10.1103/PhysRevLett.111.121801}{\emph{Phys. Rev.
  Lett.} {\bfseries 111} (2013) 121801},
  [\href{https://arxiv.org/abs/1305.6837}{{\ttfamily 1305.6837}}].

\bibitem{Alioli:2016fum}
S.~Alioli et~al., \emph{{Precision studies of observables in $p p \rightarrow W
  \rightarrow l\nu _l$ and $ pp \rightarrow \gamma ,Z \rightarrow l^+ l^-$
  processes at the LHC}},
  \href{https://doi.org/10.1140/epjc/s10052-017-4832-7}{\emph{Eur. Phys. J. C}
  {\bfseries 77} (2017) 280},
  [\href{https://arxiv.org/abs/1606.02330}{{\ttfamily 1606.02330}}].

\bibitem{Denner:2019vbn}
A.~Denner and S.~Dittmaier, \emph{{Electroweak Radiative Corrections for
  Collider Physics}},
  \href{https://doi.org/10.1016/j.physrep.2020.04.001}{\emph{Phys. Rept.}
  {\bfseries 864} (2020) 1--163},
  [\href{https://arxiv.org/abs/1912.06823}{{\ttfamily 1912.06823}}].

\bibitem{Franceschini:2017xkh}
R.~Franceschini, G.~Panico, A.~Pomarol, F.~Riva and A.~Wulzer,
  \emph{{Electroweak Precision Tests in High-Energy Diboson Processes}},
  \href{https://doi.org/10.1007/JHEP02(2018)111}{\emph{JHEP} {\bfseries 02}
  (2018) 111}, [\href{https://arxiv.org/abs/1712.01310}{{\ttfamily
  1712.01310}}].

\bibitem{Panico:2017frx}
G.~Panico, F.~Riva and A.~Wulzer, \emph{{Diboson interference resurrection}},
  \href{https://doi.org/10.1016/j.physletb.2017.11.068}{\emph{Phys. Lett. B}
  {\bfseries 776} (2018) 473--480},
  [\href{https://arxiv.org/abs/1708.07823}{{\ttfamily 1708.07823}}].

\bibitem{McCullough:2023szd}
M.~McCullough, M.~Riembau and L.~Ricci, \emph{{Boundaries of universal
  theories}}, \href{https://doi.org/10.1103/PhysRevD.111.015031}{\emph{Phys.
  Rev. D} {\bfseries 111} (2025) 015031},
  [\href{https://arxiv.org/abs/2312.03834}{{\ttfamily 2312.03834}}].

\bibitem{Englert:2019zmt}
C.~Englert, G.~F. Giudice, A.~Greljo and M.~Mccullough, \emph{{The
  $\hat{H}$-Parameter: An Oblique Higgs View}},
  \href{https://doi.org/10.1007/JHEP09(2019)041}{\emph{JHEP} {\bfseries 09}
  (2019) 041}, [\href{https://arxiv.org/abs/1903.07725}{{\ttfamily
  1903.07725}}].

\bibitem{CMS:2022krd}
{\scshape CMS} collaboration, A.~Tumasyan et~al., \emph{{Search for new physics
  in the lepton plus missing transverse momentum final state in proton-proton
  collisions at $\sqrt{s} =$ 13 TeV}},
  \href{https://doi.org/10.1007/JHEP07(2022)067}{\emph{JHEP} {\bfseries 07}
  (2022) 067}, [\href{https://arxiv.org/abs/2202.06075}{{\ttfamily
  2202.06075}}].

\bibitem{ATLAS:2019lsy}
{\scshape ATLAS} collaboration, G.~Aad et~al., \emph{{Search for a heavy
  charged boson in events with a charged lepton and missing transverse momentum
  from $pp$ collisions at $\sqrt{s} = 13$ TeV with the ATLAS detector}},
  \href{https://doi.org/10.1103/PhysRevD.100.052013}{\emph{Phys. Rev. D}
  {\bfseries 100} (2019) 052013},
  [\href{https://arxiv.org/abs/1906.05609}{{\ttfamily 1906.05609}}].

\bibitem{Boos:2006xe}
E.~Boos, V.~Bunichev, L.~Dudko and M.~Perfilov, \emph{{Interference between
  $W^\prime$ and $W$ in single-top quark production processes}},
  \href{https://doi.org/10.1016/j.physletb.2007.03.064}{\emph{Phys. Lett. B}
  {\bfseries 655} (2007) 245--250},
  [\href{https://arxiv.org/abs/hep-ph/0610080}{{\ttfamily hep-ph/0610080}}].

\bibitem{Rizzo:2007xs}
T.~G. Rizzo, \emph{{The Determination of the Helicity of $W'$ Boson Couplings
  at the LHC}},
  \href{https://doi.org/10.1088/1126-6708/2007/05/037}{\emph{JHEP} {\bfseries
  05} (2007) 037}, [\href{https://arxiv.org/abs/0704.0235}{{\ttfamily
  0704.0235}}].

\bibitem{Papaefstathiou:2009sr}
A.~Papaefstathiou and O.~Latunde-Dada, \emph{{NLO production of $W$ ' bosons at
  hadron colliders using the MC@NLO and POWHEG methods}},
  \href{https://doi.org/10.1088/1126-6708/2009/07/044}{\emph{JHEP} {\bfseries
  07} (2009) 044}, [\href{https://arxiv.org/abs/0901.3685}{{\ttfamily
  0901.3685}}].

\bibitem{Accomando:2011eu}
E.~Accomando, D.~Becciolini, S.~De~Curtis, D.~Dominici, L.~Fedeli and
  C.~Shepherd-Themistocleous, \emph{{Interference effects in heavy W'-boson
  searches at the LHC}},
  \href{https://doi.org/10.1103/PhysRevD.85.115017}{\emph{Phys. Rev. D}
  {\bfseries 85} (2012) 115017},
  [\href{https://arxiv.org/abs/1110.0713}{{\ttfamily 1110.0713}}].

\bibitem{FileviezPerez:2010gw}
P.~Fileviez~Perez and M.~B. Wise, \emph{{Baryon and lepton number as local
  gauge symmetries}},
  \href{https://doi.org/10.1103/PhysRevD.82.079901}{\emph{Phys. Rev. D}
  {\bfseries 82} (2010) 011901},
  [\href{https://arxiv.org/abs/1002.1754}{{\ttfamily 1002.1754}}].

\bibitem{Schwaller:2013hqa}
P.~Schwaller, T.~M.~P. Tait and R.~Vega-Morales, \emph{{Dark Matter and
  Vectorlike Leptons from Gauged Lepton Number}},
  \href{https://doi.org/10.1103/PhysRevD.88.035001}{\emph{Phys. Rev. D}
  {\bfseries 88} (2013) 035001},
  [\href{https://arxiv.org/abs/1305.1108}{{\ttfamily 1305.1108}}].

\bibitem{GonzalezSuarez:2024dsp}
R.~Gonzalez~Suarez, B.~Pattnaik and J.~Zurita, \emph{{Leptophilic Z' bosons at
  the FCC-ee: Discovery opportunities}},
  \href{https://doi.org/10.1103/PhysRevD.111.035029}{\emph{Phys. Rev. D}
  {\bfseries 111} (2025) 035029},
  [\href{https://arxiv.org/abs/2410.12903}{{\ttfamily 2410.12903}}].

\bibitem{ALEPH:2013dgf}
{\scshape ALEPH, DELPHI, L3, OPAL, LEP Electroweak} collaboration, S.~Schael
  et~al., \emph{{Electroweak Measurements in Electron-Positron Collisions at
  W-Boson-Pair Energies at LEP}},
  \href{https://doi.org/10.1016/j.physrep.2013.07.004}{\emph{Phys. Rept.}
  {\bfseries 532} (2013) 119--244},
  [\href{https://arxiv.org/abs/1302.3415}{{\ttfamily 1302.3415}}].

\bibitem{Greljo:2024ytg}
A.~Greljo, H.~Tiblom and A.~Valenti, \emph{{New physics through flavor tagging
  at FCC-ee}},
  \href{https://doi.org/10.21468/SciPostPhys.18.5.152}{\emph{SciPost Phys.}
  {\bfseries 18} (2025) 152},
  [\href{https://arxiv.org/abs/2411.02485}{{\ttfamily 2411.02485}}].

\bibitem{Garosi:2023bvq}
F.~Garosi, D.~Marzocca and S.~Trifinopoulos, \emph{{LePDF: Standard Model PDFs
  for high-energy lepton colliders}},
  \href{https://doi.org/10.1007/JHEP09(2023)107}{\emph{JHEP} {\bfseries 09}
  (2023) 107}, [\href{https://arxiv.org/abs/2303.16964}{{\ttfamily
  2303.16964}}].

\bibitem{Banks:2020gpu}
H.~Banks and M.~McCullough, \emph{{Charting the Fifth Force Landscape}},
  \href{https://doi.org/10.1103/PhysRevD.103.075018}{\emph{Phys. Rev. D}
  {\bfseries 103} (2021) 075018},
  [\href{https://arxiv.org/abs/2009.12399}{{\ttfamily 2009.12399}}].

\end{thebibliography}\endgroup

\end{document}